\documentclass[aps,showpacs,preprintnumbers,amsmath,amssymb,twocolumn]{revtex4}
\preprint{Charit\'e, Biochemistry 03/01}
\usepackage{graphicx}

\usepackage{graphics}
\begin{document}
\bibliographystyle{apsrev}
\title{Hydrodynamics and transport coefficients for Granular Gases}
\author{Nikolai Brilliantov$^{1,2}$, Thorsten P\"{o}schel$^1$}
\affiliation{$^1$Institut f\"ur Biochemie, Charit\'e, Monbijoustra{\ss}e 2, 10117 Berlin, Germany\\
$^2$Moscow State University, 119899 Moscow, Russia}
\date{\today}

\begin{abstract}
The hydrodynamics of granular gases of viscoelastic particles, whose collision is described by an impact-velocity dependent coefficient of restitution, is developed using a modified Chapman-Enskog approach. We derive the hydrodynamic equations and the according transport coefficients with the assumption that the shape of the velocity distribution function follows adiabatically the decaying temperature. We show numerically that this approximation is justified up to intermediate dissipation. The transport coefficients and the coefficient of cooling are expressed in terms of the elastic and dissipative parameters of the particle material and by the gas parameters. The dependence of these coefficients on temperature differs qualitatively from that obtained with the simplifying assumption of a constant coefficient of restitution which was used in previous studies. The approach formulated for gases of viscoelastic particles may be applied also for other impact-velocity dependencies of the restitution coefficient. 

\end{abstract}


\maketitle
\section{Introduction} 
Granular systems composed of a large number of dissipatively interacting particles behave in many respects as a continuous medium and may be, in principle, described by a set of hydrodynamic equations with appropriate boundary conditions. Although this approach is successively used in various fields of engineering and soil mechanics (e.g. \cite{Kolymbas2000,Vermeer2001}) a first-principle theory for dense granular media is still lacking.
Hydrodynamics may be also applied to much simpler systems, such as rarefied granular gases. It has been used to describe many different processes, e.g. rapid granular flows, structure formation, etc. (see \cite{PoeschelLuding:2001} for an overview). 
For these systems the hydrodynamic equations are not postulated but derived from the Boltzmann equation. The corresponding transport coefficients are not phenomenological constants, instead they are obtained by regular methods, such as the Grad method \cite{Grad:1949} or the Chapman-Enskog method \cite{ChapmanCowling:1970}. In most of the studies, which address the derivation of hydrodynamic equations and kinetic coefficients, it was assumed that the coefficient of restitution $\varepsilon$ is a material constant, e.g. \cite{LunSavageJeffreyChepurniy:1984,JenkinsRichman:1985,GoldshteinShapiro:1995,SelaGoldhirsch:1998,BreyDuftyKimSantos:1998,BreyRuizMonteroCubero:1999,GarzoDufty:1999,BreyCubero:2000}. 

This assumption simplifies the analysis enormously, however, it is neither in agreement with experimental observations (e.g. \cite{Goldsmit1960,BridgesHatzesLin:1984,KuwabaraKono:1987}) nor with basic mechanics of particle collisions \cite{Ramirez:1999}. The coefficient of restitution depends on the impact velocity $g$ and tends to unity for very small $g$. As a consequence, particles behave more and more elastically as the average velocity of the grains decreases. The simplest collision model which accounts for dissipative material deformation, is the model of viscoelastic particles. It is assumed in this model that the elastic stress in the bulk of the particle material depends linearly on the strain, whereas the dissipative stress depends linearly on the strain rate. 

Based on the fundamental work by Hertz \cite{Hertz:1882} the interaction force between colliding viscoelastic spheres has been derived \cite{BSHP}. Using this generalized Hertz law the coefficient of restitution of viscoelastic particles can be given as a function of the impact velocity and material parameters \cite{SchwagerPoeschel:1998}:  
\begin{equation}
\varepsilon=1- C_1 A \alpha^{2/5} g ^{1/5} +\frac35 C_1^2 A^2  \alpha^{4/5} g^{2/5}  \mp
 \dots
\label{epsilon}
\end{equation}
with
\begin{equation}
  \label{eq:gdef}
  g\equiv \left| \vec{e} \cdot \left(\vec{v}_{1}-\vec{v}_{2}\right) \right| 
     = \left| \vec{e} \cdot \vec{v}_{12} \right| \, .
\end{equation}
The unit vector $\vec{e}=\vec{r}_{12}/r_{12}$ specifies the collision geometry, i.e., the relative position $\vec{r}_{12}=\vec{r}_1 - \vec{r}_2$ of the particles at the collision instant. Their pre-collision velocities are given by $\vec{v}_1$ and $\vec{v}_2$. The elastic constant 
\begin{equation}
\alpha= \left( \frac32 \right)^{3/2}
\frac{ Y\sqrt{R^{\,\mbox{\footnotesize eff}}}}{
    m^{\mbox{\footnotesize eff}}\left( 1-\nu ^2\right)
}\, ,
\end{equation}
depends on the effective mass and radius $m^{\rm eff}\equiv m_1m_2/(m_1+m_2)$, $R^{\rm eff}\equiv R_1R_2/(R_1+R_2)$ of the colliding spheres, on the Young modulus $Y$, and on the Poisson ratio $\nu$ of the particle material. The dissipative coefficient $A$ is a function of dissipative and elastic constants (see \cite{BSHP} for details). Finally, $C_1$ is a numerical constant \cite{SchwagerPoeschel:1998,Ramirez:1999}:
\begin{equation}
\label{eq:coeffepsC1}
    C_1=\frac{\sqrt{\pi}}{2^{1/5}5^{2/5}} \frac{\Gamma \left( 3/5 \right)}{\Gamma\left(21/10\right)}
       \approx 1.15344 \, .
\end{equation}
Equation \eqref{epsilon} describes pure viscoelastic interaction. The assumption of viscoelastic deformation is justified if the impact velocity is not too large to avoid plastic deformation of the particles and not too small to neglect surface effects such as adhesion, van der Waals forces etc. We also assume that the rotational degrees of freedom of particles may be neglected and consider a granular gas of identical particles in the absence of external forces.  Then the coefficient of restitution gives the velocities of particles after a collision $\vec{v}_1^{\, \prime}$, $\vec{v}_2^{\, \prime}$ in terms of their values before the collision: 
\begin{equation}
\label{eq:v1v2prime}
\vec{v}_{1/2}^{\, \prime} = \vec{v}_{1/2}\mp \frac{1+\varepsilon}{2}
\left( \vec{e} \cdot \vec{v}_{12} \right)\vec{e} 
\end{equation}

The impact velocity dependence of the coefficient of restitution implies serious consequences for the granular gas dynamics: For the simplified case $\varepsilon=\mbox{const.}$ the form of the velocity distribution function $f(\vec{v},t)$  is characterized by a time-independent scaled function $\tilde{f} \left(\vec{c}\right)$. It depends only on the scaled velocity $\vec{c}=\vec{v}/v_T$, where $v_T(t) = \sqrt{2T(t)/m}$ is the thermal velocity and $T(t)$ is the granular temperature \cite{EsipovPoeschel:1995,NoijeErnst:1998,BrilliantovPoeschelStability:2000}. The distribution function depends on time only via the time dependence of the granular temperature, its shape is time independent. The small deviations of the velocity distribution function from the Maxwell distribution are determined by the time-independent coefficient of restitution. For granular gases of viscoelastic particles, however, the effective value of $\varepsilon$ changes with time along with the thermal velocity $v_T(t)$, which gives a typical velocity of the particles. Therefore, the shape of the velocity distribution function evolves in a rather complicated way \cite{BrilliantovPoeschel:2000visc}. 

The time dependence of the velocity distribution function has to be taken into account when the hydrodynamic equations and the transport coefficients are derived. Since the simple scaling is violated, the standard methods of kinetic theory of gases, developed for $\varepsilon=\mbox{const.}$ (e.g. \cite{BreyDuftyKimSantos:1998,GarzoDufty:1999,RamirezRissoSoto:2000}) must be revised. 

As it follows from our analysis  the transport coefficients  for gases of viscoelastic particles depend on  temperature and time rather differently as compared with the case $\varepsilon = \mbox{const.}$ Correspondingly, the behavior of gases of viscoelastic particles differs qualitatively from that of gases of particles with the simplified collision model  $\varepsilon=\mbox{const.}$

We wish to remark that the impact velocity dependence of $\varepsilon$  has been already taken into account in Ref. \cite{LunSavage1986} for the hydrodynamic description of granular shear flow. In this study an empirical expression of the coefficient of restitution was applied and a Maxwellian velocity distribution was assumed. Moreover, the authors have used the standard Chapman-Enskog method without the modifications  required for gases of dissipatively colliding particles. These modifications for the case of $\varepsilon = {\rm const.}$ have been extensively elaborated in \cite{BreyDuftyKimSantos:1998}.

The aim  of the present study is to develop  a continuum description of granular gases of viscoelastic particles. We derive the hydrodynamic equations along with the transport coefficients and the coefficient of cooling. In the rest of the paper, in Sec. II-V, we discuss in detail the most important case of the three-dimensional gases, while  in Sec. VI we present the results for the two-dimensional systems, which are frequently addressed in molecular dynamics studies.

\section{Velocity distribution and temperature in the homogeneous cooling state}
\subsection{Evolution equations for temperature and for the second Sonine coefficient}
The Boltzmann equation for a granular gas of viscoelastic particles in the homogeneous cooling state reads  \cite{BrilliantovPoeschel:2000visc,BrilliantovPoeschelGG} 
\begin{eqnarray}
\label{collint1}
\frac{\partial}{\partial t}f\left(\vec{v}_1,t\right)  &=&\sigma^2 g_2(\sigma) \int d \vec{v}_2 \int d\vec{e} \, 
\Theta\left( -\vec{v}_{12} \cdot \vec{e}\,\right) \left|\vec{v}_{12} \cdot \vec{e}\,\right| \nonumber \\
&&\times \left[ \chi f \left(\vec{v}^{\,\prime \prime}_1,t \right)
f \left(\vec{v}^{\,\prime \prime}_2,t \right) -
f \left(\vec{v}_1,t \right)
f \left(\vec{v}_2,t \right) \right] \nonumber \\
&\equiv& g_2(\sigma) I \left( f,f \right) \, ,  
\end{eqnarray}
where $\sigma=2R$ is the particle diameter. The velocities $\vec{v}^{\,\prime \prime}_1$ and  $\vec{v}^{\,\prime \prime}_2 $ denote the pre-collision velocities of the inverse collision, which leads to the after-collision velocities $\vec{v}_1$ and  $\vec{v}_2 $. The factor $\left|\vec{v}_{12} \cdot \vec{e}\,\right|$ characterizes the length of the collision cylinder of cross-section $\sigma^2$ and the Heaviside step-function $\Theta\left( -\vec{v}_{12} \cdot \vec{e}\right)$ assures that only approaching particles collide. The contact value of pair correlation function $g_2(\sigma)$ accounts for the increased collision frequency due to excluded volume effects. Finally, the factor $\chi$ in the gain term accounts for the Jacobian of the transformation $\left(\vec{v}^{\,\prime \prime}_1, \vec{v}^{\,\prime \prime}_2\right) \to \left(\vec{v}_1, \vec{v}_2\right) $ and for the ratio of the lengths of the collision cylinders $\left|\vec{v}^{\,\prime \prime}_{12} \cdot \vec{e}\,\right| /  \left|\vec{v}_{12} \cdot \vec{e}\,\right|$ for the direct and the inverse collision. 
For the case of spheres colliding  with a constant coefficient of restitution $\chi = 1/\varepsilon^2$, while for viscoelastic spheres, with $\varepsilon=\varepsilon(g)$ given by  Eq. \eqref{epsilon}, it reads \cite{BrilliantovPoeschel:2000visc,BrilliantovPoeschelGG}
\begin{multline}
\label{CHI}
\chi = 1 + \frac{11}{5}C_1 A \alpha^{2/5} \left|\vec{v}_{12} \cdot \vec{e}\,\right|^{1/5} \\ 
+ \frac{66}{25}C_1^2 A^2 \alpha^{4/5} \left|\vec{v}_{12} \cdot \vec{e}\, \right|^{2/5} +\dots \, .
\end{multline}
The dependence of $\chi$ on the impact velocity does not allow to derive from the Boltzmann equation a time-independent equation for the scaled distribution function $\tilde{f}$. Contrary to the case of $\varepsilon = \mbox{const.}$ \cite{EsipovPoeschel:1995,NoijeErnst:1998,BrilliantovPoeschelStability:2000} the scaled distribution function depends  explicitly on time. Therefore, we write for a gas of viscoelastic particles
\begin{equation}
\label{1genscal}
f\left(\vec{v}, t\right)=\frac{n}{v_T^3(t)}\tilde{f}(\vec{c}, t)  
\qquad \vec{c}=\frac{\vec{v}}{v_T(t)}  \,,
\end{equation}
with the number density of the granular gas $n$ and the thermal velocity $v_T(t)$ defined by the granular temperature:
\begin{equation}
\label{deftemp1}
\frac{3}{2} n T(t)= \int d \vec{v}\, \frac{mv^2}{2}
f(\vec{v},t)
=\frac32 n \,\,  \frac{mv_T^2(t)}{2} \,.
\end{equation}
Hence, the shape of the velocity distribution function, characterized by the rescaled function $\tilde{f}(\vec{c}, t)$, does not persist but evolves along with temperature \cite{BrilliantovPoeschel:2000visc,BrilliantovPoeschelGG}. We wish to stress that the time dependence of $\tilde{f}(\vec{c}, t)$ is caused by the dependence of the factor $\chi$ on the impact velocity. Contrary, for a gas of simplified particles ($\varepsilon=\mbox{const.}$) we obtain $\chi=1/\varepsilon^2=\mbox{const.}$ and, therefore, the rescaled distribution function is time independent,  $\tilde{f}(\vec{c}, t)= \tilde{f}(\vec{c}\,)$.

For slightly dissipative particles the velocity distribution function is close to the Maxwell distribution. It may be described by a Sonine polynomial expansion \cite{GoldshteinShapiro:1995,NoijeErnst:1998,HuthmannOrzaBrito:2000,BrilliantovPoeschelStability:2000}:
\begin{equation}
\label{genSoninexp}
\tilde{f}(\vec{c}, t)
=\phi(c) \left( 1 + \sum_{p=1}^{\infty} a_p(t) S_p\left(c^2\right) \right) \,,
\end{equation}
where $\phi(c)\equiv \pi^{-3/2}\exp\left(-c^2\right)$ is the scaled Maxwell distribution, $S_p(x)$ are the Sonine polynomials, 
\begin{equation}
  \begin{split}
S_0(x)&=1 \\
S_1(x)&=-x^2+\frac{3}{2} \\
S_2(x)&=\frac{x^2}{2}-\frac{5x}{2}+\frac{15}{8}\,,~~~ \mbox{etc.}     
  \end{split}
\end{equation}
and $a_k(t)$ are the {\em time-dependent} Sonine coefficients, which characterize the form of the velocity distribution \cite{BrilliantovPoeschel:2000visc,BrilliantovPoeschelStability:2000}. The first Sonine coefficient is trivial, $a_1=0$,  due to the definition of temperature \cite{NoijeErnst:1998,BrilliantovPoeschelStability:2000}, while the other coefficients quantify deviations of the moments $\left< c^k \right>$ of the velocity distribution from the moments of the Maxwell distribution $\left< c^k \right>_0$, e.g. 
\begin{equation}
    a_2 = \frac{ \left<c^4\right> - \left<c^4\right>_0 }{\left<c^4\right>_0 }\, . 
\label{eq:SonineMoments}
\end{equation}
Thus, the first nontrivial Sonine coefficient $a_2$ characterizes the fourth moment of the distribution function. 

For small enough inelasticity ($\varepsilon \gtrsim 0.6$) the distribution function is well approximated by the second Sonine coefficient $a_2$ \cite{NoijeErnst:1998,HuthmannOrzaBrito:2000,BrilliantovPoeschelStability:2000}, i.e. higher coefficients $a_k = 0$ for $k \ge 3$ may be neglected. With this approximation the evolution a granular gas of viscoelastic particles in the homogeneous cooling state is described by a set of coupled equations for the granular temperature and for the second Sonine coefficient \cite{BrilliantovPoeschel:2000visc,BrilliantovPoeschelGG}:
\begin{eqnarray}
\label{dTdt}
\frac{d T}{d t} &=& -\frac23 BT\mu_2 \equiv - \zeta T\\
\label{eqa2}
\frac{d a_2}{d t}&=& \frac43\, B\mu_2 \left(1+a_2 \right)-\frac{4}{15} \, B\mu_4 \,.
\end{eqnarray}
The coefficient $B\equiv B(t)=v_T(t)\sigma^2 g_2(\sigma) n$ is proportional to the mean collision frequency. The moments of the collision integral,  $\mu_2$ and $\mu_4$ read with the approximation $\tilde{f}= \phi (c) \left[ 1+a_2(t)S_2\left(c^2\right)\right]$:
\begin{multline}
\label{mupa2}
\mu_p=-\frac12 \int d\vec{c}_1\int d \vec{c}_2 \int d\vec{e} \, \Theta\left(-\vec{c}_{12} \cdot \vec{e}\,\right) \left|\vec{c}_{12} \cdot \vec{e}\,\right| \phi\left(c_1\right) \\[0.2cm]
\times \phi\left(c_2\right) \left\{1+a_2\left[S_2\left(c_1^2\right)+S_2\left(c_2^2\right) \right] + a_2^2\,S_2\left(c_1^2\right)S_2\left(c_2^2\right) \right\}\\[0.2cm]
\times  \Delta \left(c_1^{\,p}+c_2^{\,p}\right) 
\end{multline}
where 
\begin{equation}
\Delta \psi\left(\vec{c}_i\right) \equiv \psi(\vec{c}_i^{\,\prime})-\psi(\vec{c}_i)  
\end{equation}
denotes the change of some function $\psi\left( \vec{c}_i\right)$ according to a collision. The coefficients $\mu_p$ depend on time via the time dependence of $a_2$. For small enough dissipation the moments $\mu_2$ and $\mu_4$ may be obtained as expansions in the time-dependent dissipative parameter $\delta^{\, \prime}$, 
\begin{equation}
\label{deltaprime}
\delta^{\, \prime} (t) \equiv A \alpha^{2/5} \left[2T(t)\right]^{1/10} \equiv \delta \left[\frac{2T(t)}{T_0} \right]^{1/10} \, , 
\end{equation}
with $\delta \equiv A \alpha^{2/5}T_0^{1/10}$ and with  the initial temperature $T_0$. These expansions read \cite{BrilliantovPoeschel:2000visc,BrilliantovPoeschelGG}
\begin{equation}
\label{mu2A}
\begin{split}
\mu_2&= \sum_{k=0}^{2} \sum_{n=0}^2 {\cal A}_{kn}
\delta^{\, \prime \,k } a_2^n\\ 
\mu_4&= \sum_{k=0}^{2} \sum_{n=0}^2 {\cal B}_{kn}
\delta^{\, \prime \,k } a_2^n  
\end{split}
\end{equation}
where ${\cal A}_{kn}$ and ${\cal B}_{kn}$ are numerical coefficients. They may be written in the compact matrix notation (rows refer to the first index):
\begin{equation}
  \label{eq:A}
  \begin{split}
\hat{\cal A} &= \left(
{  \renewcommand{\arraystretch}{1.5}
\begin{tabular}{ccc}
    $0$         & $0$                         & $0$\\
    $\omega_0$  & $ \frac{6}{25}\omega_0 $     & $\frac{21}{2500} \omega_0 $\\
    $-\omega_1$ & $ -\frac{119}{400}\omega_1$ & $-\frac{4641}{640000}\omega_1$
  \end{tabular}  }
\right)\\[0.2cm]
\hat{\cal B} &= \left(
  {  \renewcommand{\arraystretch}{1.5}
    \begin{tabular}{ccc}
    $0$                      & $4\sqrt{2 \pi}$                  & $\frac{1}{8}\sqrt{2 \pi}$ \\
    $\frac{28}{5}\omega_0$   & $\frac{903}{125}\omega_0$        & $-\frac{567}{12500}\omega_0 $\\
    $-\frac{77}{10}\omega_1$ & $-\frac{476973}{44 000}\omega_1$ & $\frac{4459833}{704 00 000}\omega_1 $
  \end{tabular}}  
\right)
      \end{split}
\end{equation}
with 
\begin{equation}
\label{eq:omegadef}
  \begin{split}
\omega_0 &\equiv 2 \sqrt{2 \pi} 2^{1/10} \Gamma \left (\frac{21}{10} \right)\approx 6.485\\
\omega_1 &\equiv \sqrt{2\pi} 2^{1/5} \Gamma \left (\frac{16}{5} \right) C_1^2 \approx 9.285\,.     
  \end{split}
\end{equation}
The coupled equations \eqref{dTdt} and \eqref{eqa2} together with \eqref{deltaprime}, \eqref{mu2A} and \eqref{eq:A} determine the evolution of a granular gas of viscoelastic particles in the homogeneous cooling state. In particular, they define the velocity distribution function which is the starting point for the investigation of inhomogeneous gases.  

\subsection{Adiabatic approximation for the second Sonine coefficient} 

In the limit of small dissipation, $\delta \ll 1 $, the coupled equations \eqref{dTdt},\eqref{eqa2} may be solved analytically. In linear  approximation with respect to $\delta $ the solution reads \cite{BrilliantovPoeschel:2000visc,BrilliantovPoeschelGG}
\begin{equation}
\label{T(t)del1}
\frac{T(t)}{T_0}= \left(1 + \frac{t}{\tau_0} \right)^{-5/3}
\end{equation}
where we introduce the characteristic time
\begin{equation}
\label{tau0}
\tau_0^{-1}=\frac{16}5 q_0 \delta \tau_c(0)^{-1}= \frac{48}{5}q_0 \delta  \,4 \sigma^2 n\sqrt{\frac{\pi T_0}{m}} \, ,
\end{equation}
with the initial mean collision time $\tau_c(0)$ and the constant 
\begin{equation}
q_0=2^{1/5}\Gamma\left(\frac{21}{10}\right) \frac{C_1}{8} \approx 0.173\,.   
\end{equation}
Correspondingly in linear approximation the second Sonine coefficient depends on time as  \cite{BrilliantovPoeschel:2000visc}
\begin{equation}
\label{a21gensolLi}
a_2(t)=-\frac{12}{5}w(t)^{-1}
\left\{ {\rm Li} \left[ w(t) \right]-{\rm Li} \left[w(0) \right] \right\}\,,
\end{equation}
with
\begin{equation}
\label{eq:defwt}
w(t) \equiv \exp \left[ \left(q_0 \delta \right)^{-1} \left(1+\frac{t}{\tau_0} \right)^{1/6} \right]
\end{equation}
and with the logarithmic integral
\begin{equation}
{\rm Li}(x) \equiv \int_0^x \frac{1}{\ln(t)} dt\,.
\end{equation}

For small dissipation $\delta$ the time dependence of $a_2$ reveals two different regimes: (i) fast initial relaxation on the mean collision-time scale $\sim \tau_c(0)$ and (ii) subsequent slow evolution on the time scale $\sim \tau_0 \gg \tau_c(0)$, i.e. on the  time scale of the temperature evolution. Therefore the coefficient $a_2$ (and hence  the form of the velocity distribution function) evolves in accordance with temperature.

\begin{figure}[htbp]
   \centerline{\includegraphics[width=8cm,clip=]{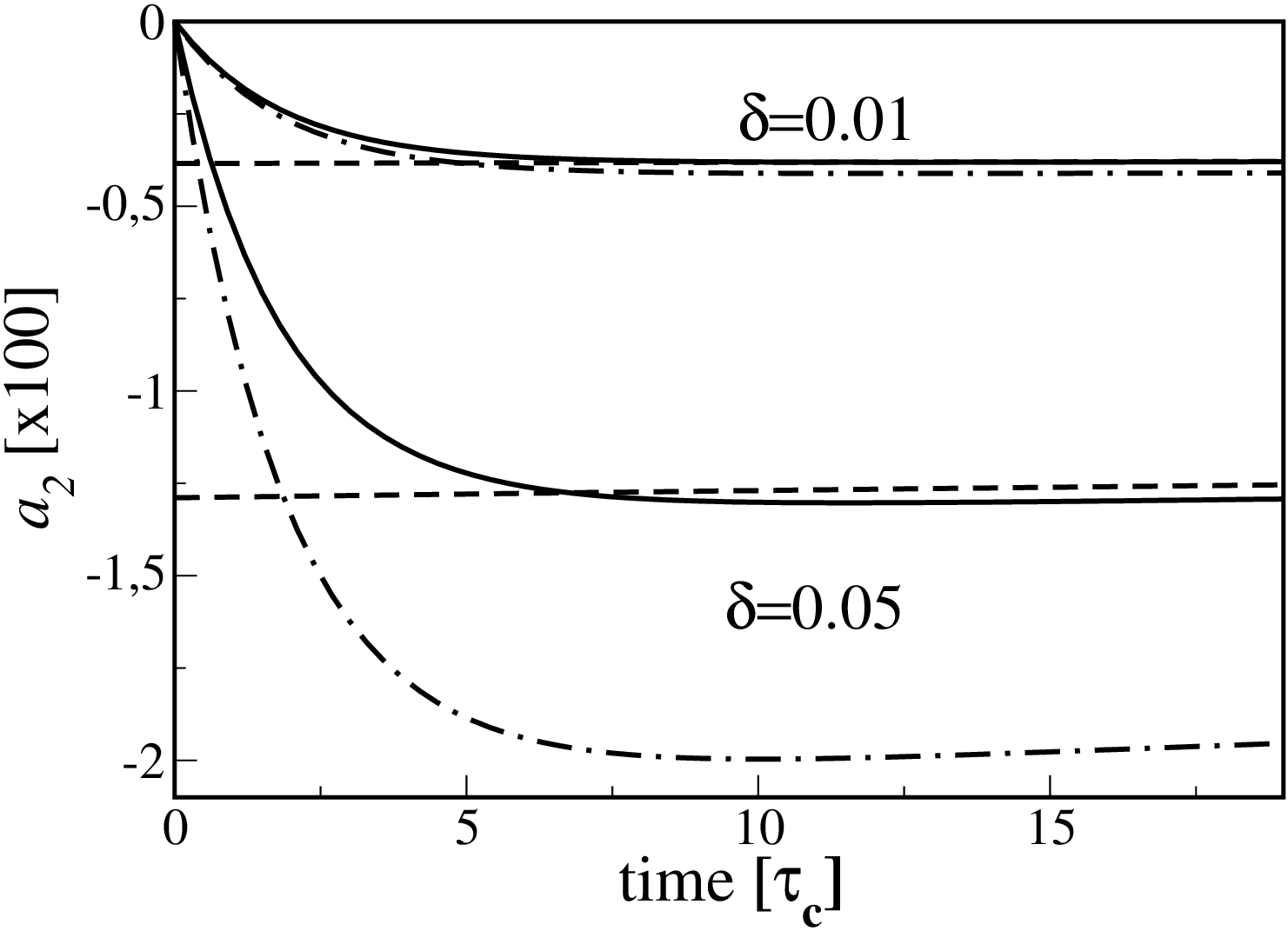}}
\vspace*{0.2cm}
   \centerline{\includegraphics[width=8cm,clip=]{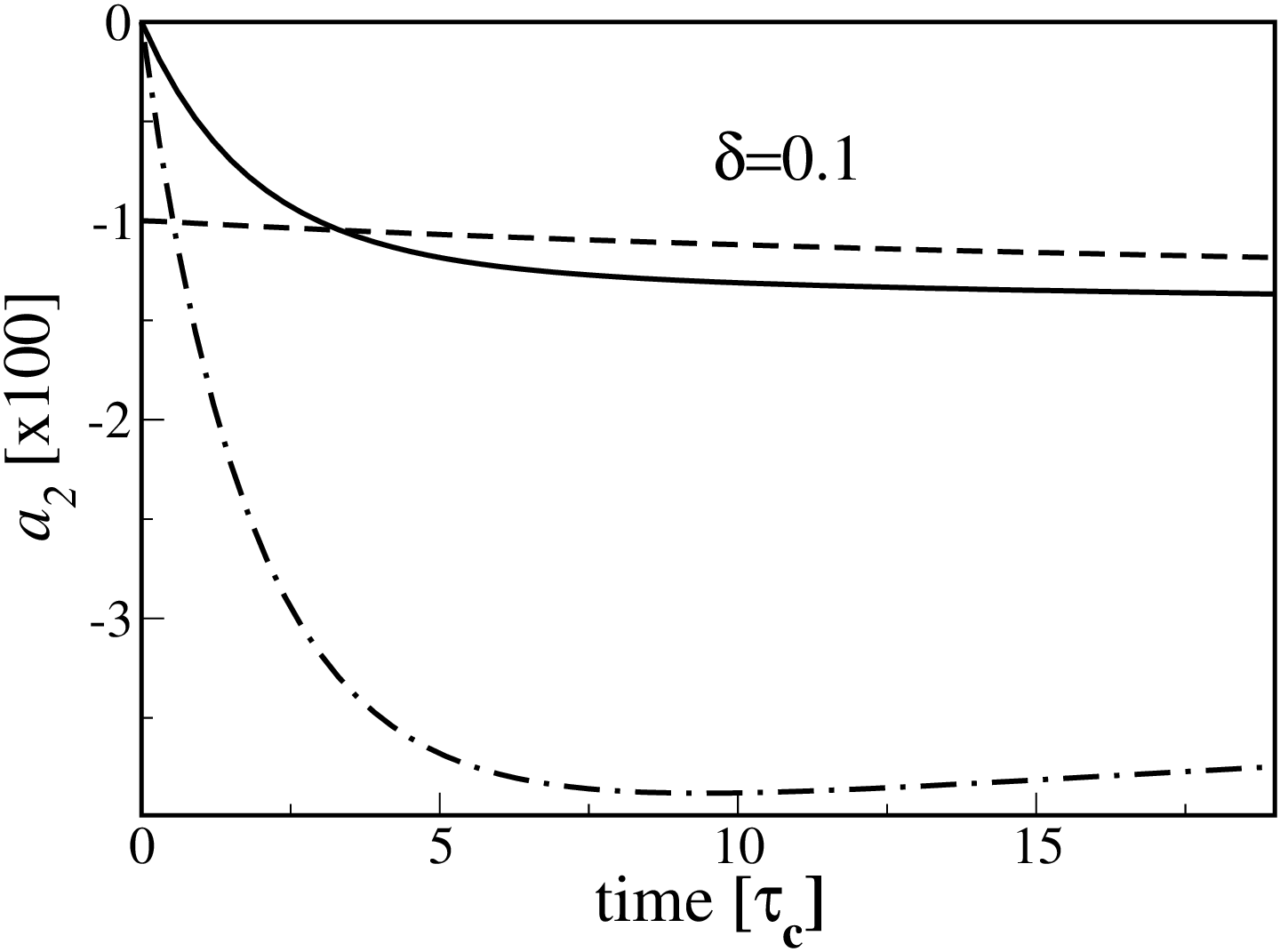}}
\vspace*{0.2cm}
   \centerline{\includegraphics[width=8cm,clip=]{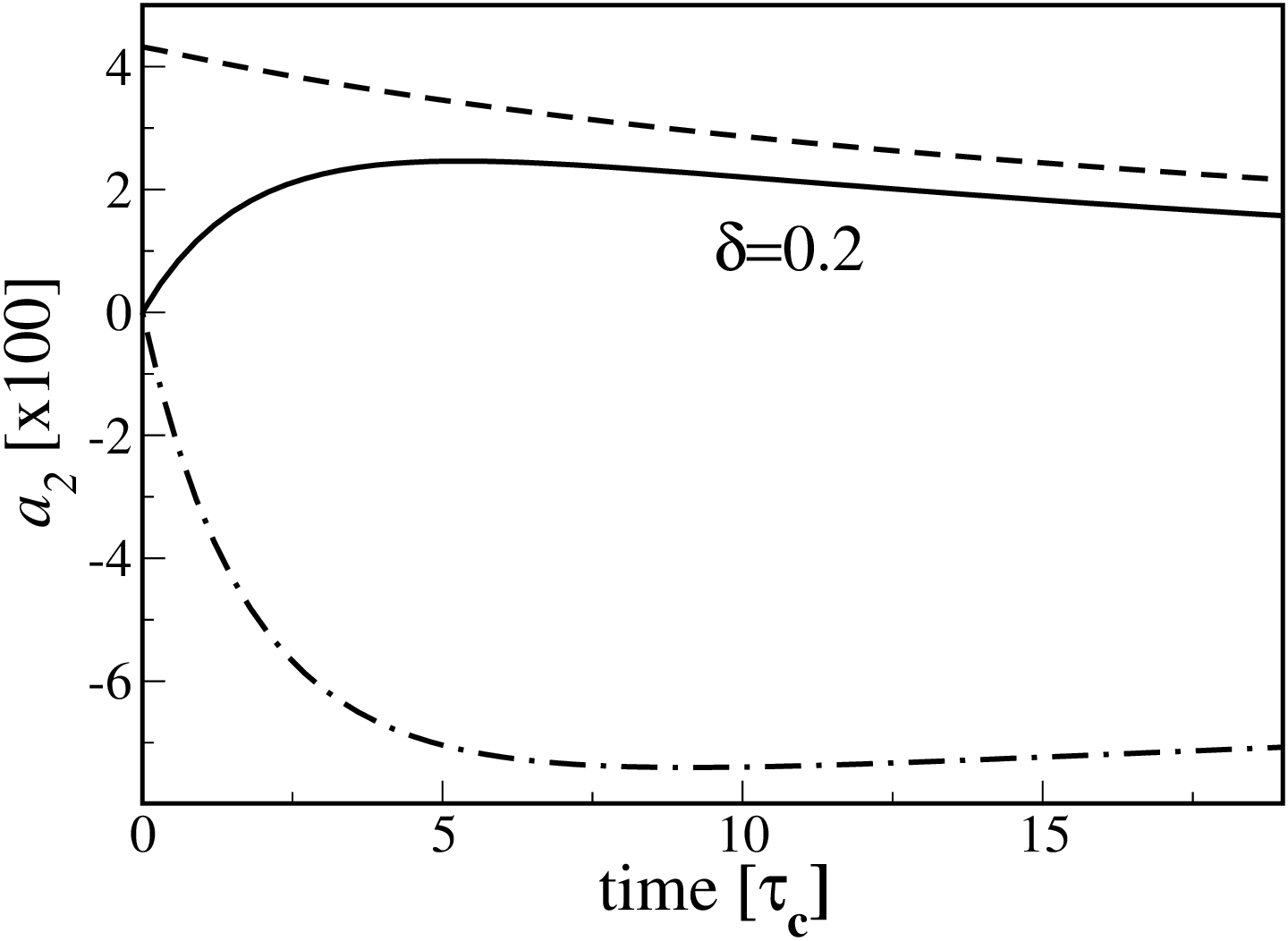}}
  \caption{Evolution of the second Sonine coefficient $a_2$ in the homogeneous cooling state on the short time scale. Solid line -- numerical solution of Eqs.(\ref{dTdt},\ref{eqa2}), dashed line -- adiabatic approximation (\ref{a2adiab}), dot-dashed line -- the result of the linear theory,  Eqs.(\ref{a21gensolLi},\ref{eq:defwt}). The time is given in collision units $\tau_c(0)$. 
}
  \label{fig:a2shortwLi}
\end{figure}

\begin{figure}[htbp]
   \centerline{\includegraphics[width=8cm,clip=]{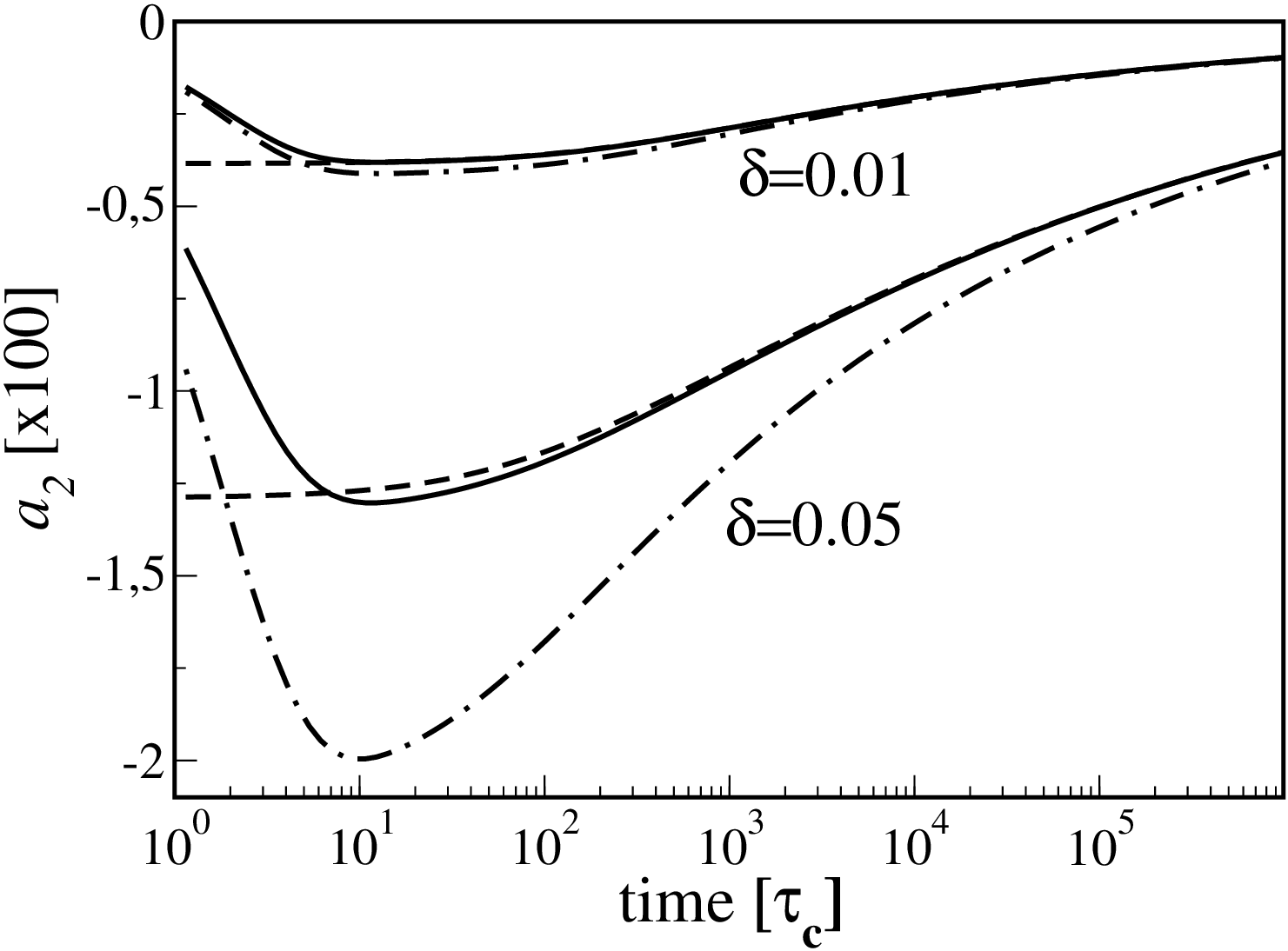}}
\vspace*{0.1cm}
   \centerline{\includegraphics[width=8cm,clip=]{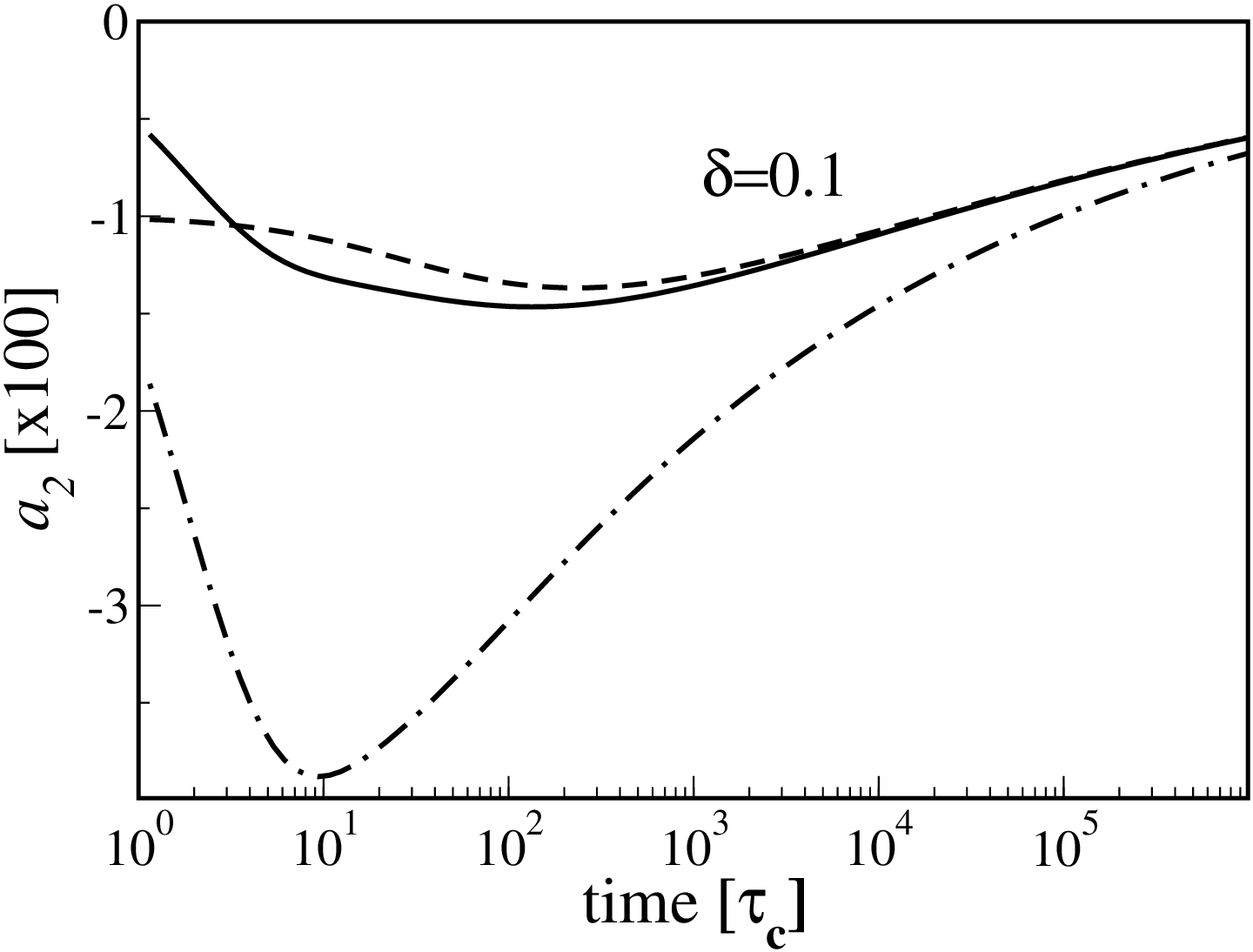}}
\vspace*{0.1cm}
   \centerline{\includegraphics[width=8cm,clip=]{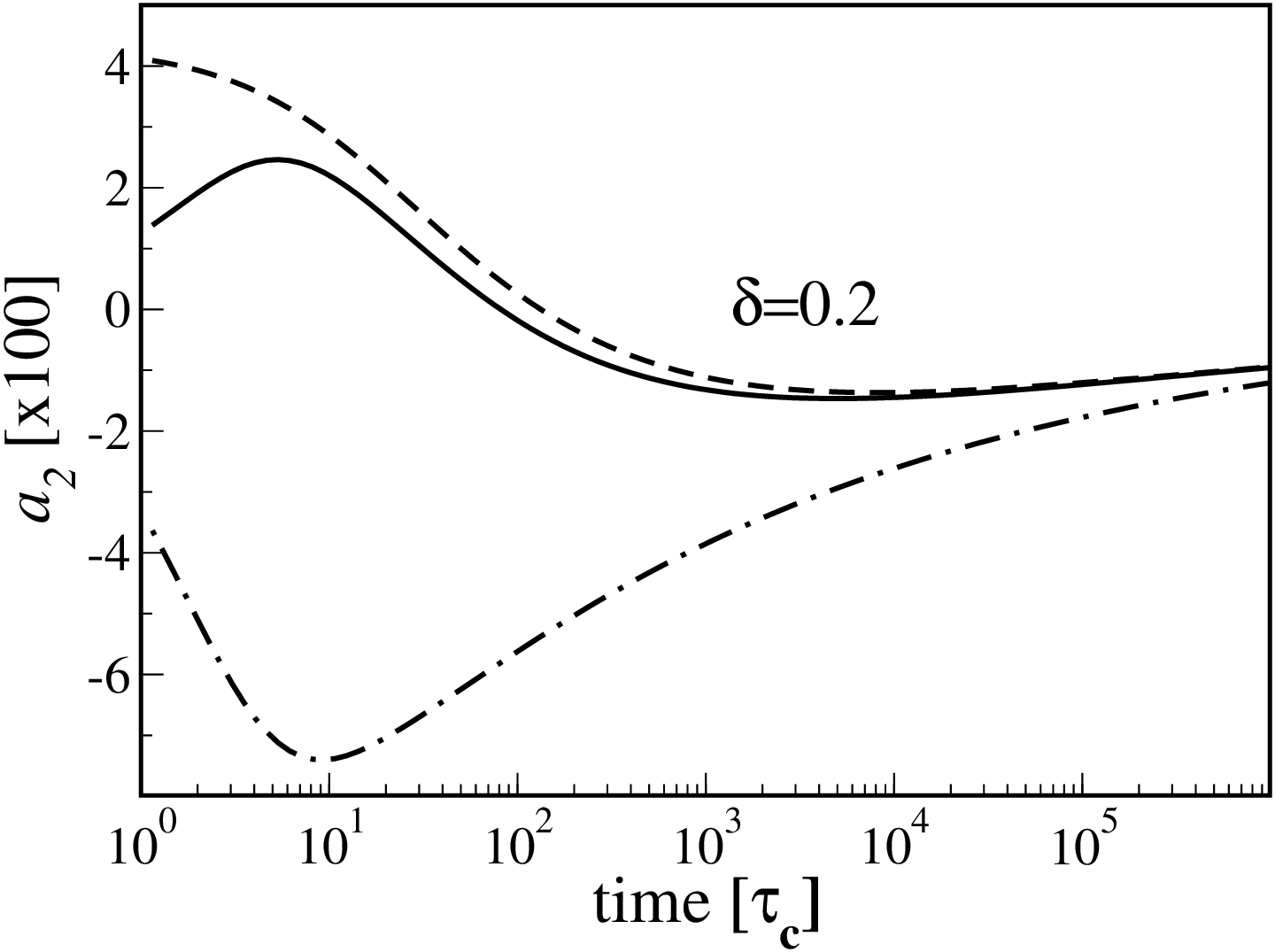}}
  \caption{The same as in Fig. \ref{fig:a2shortwLi}, but for longer time. The adiabatic approximation improves as the system evolves.}
  \label{fig:a2longwLi}
\end{figure}

\begin{figure}[htbp]
   \centerline{\includegraphics[width=8cm,clip=]{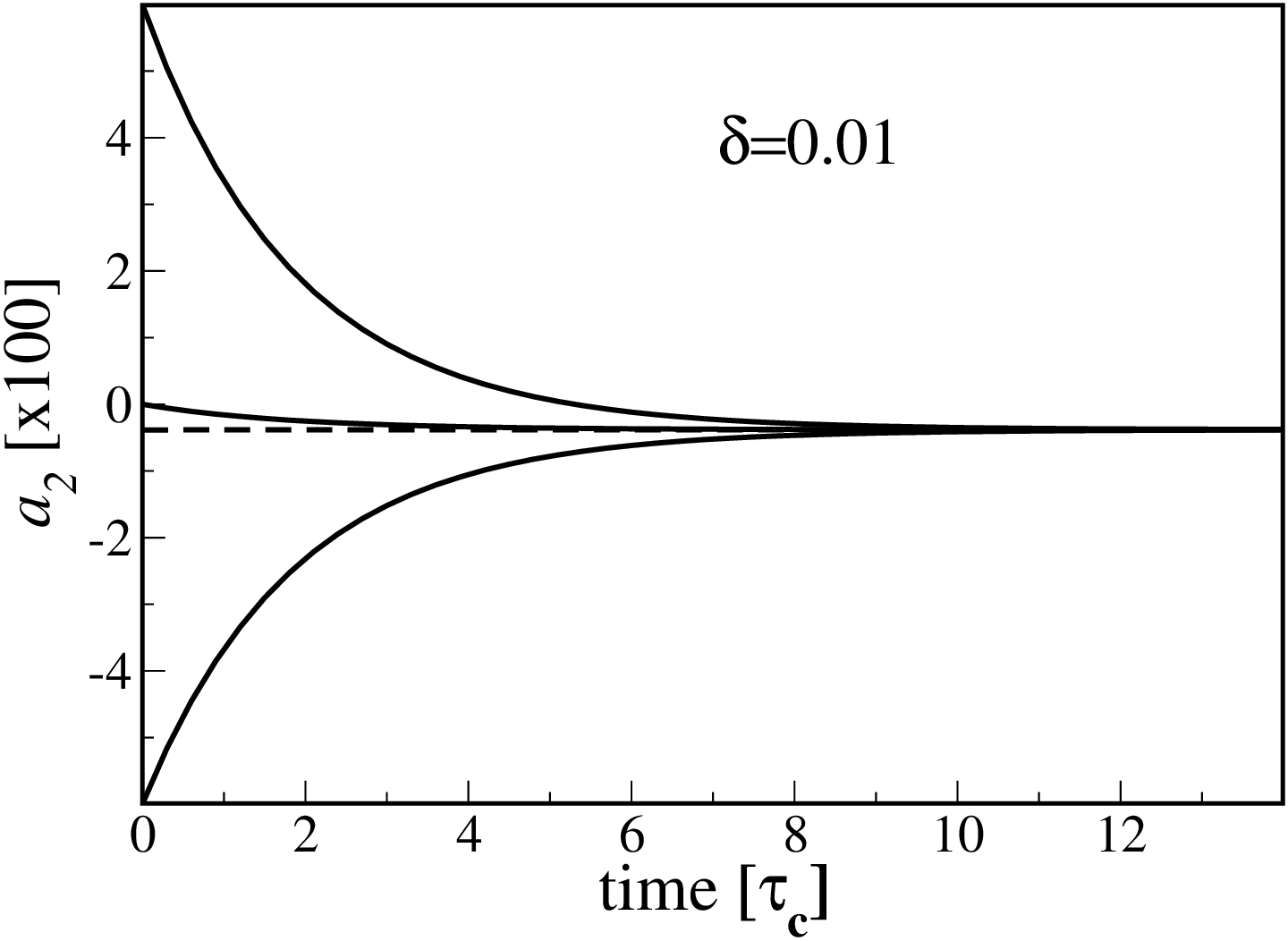}}
\vspace*{0.2cm}
   \centerline{\includegraphics[width=8cm,clip=]{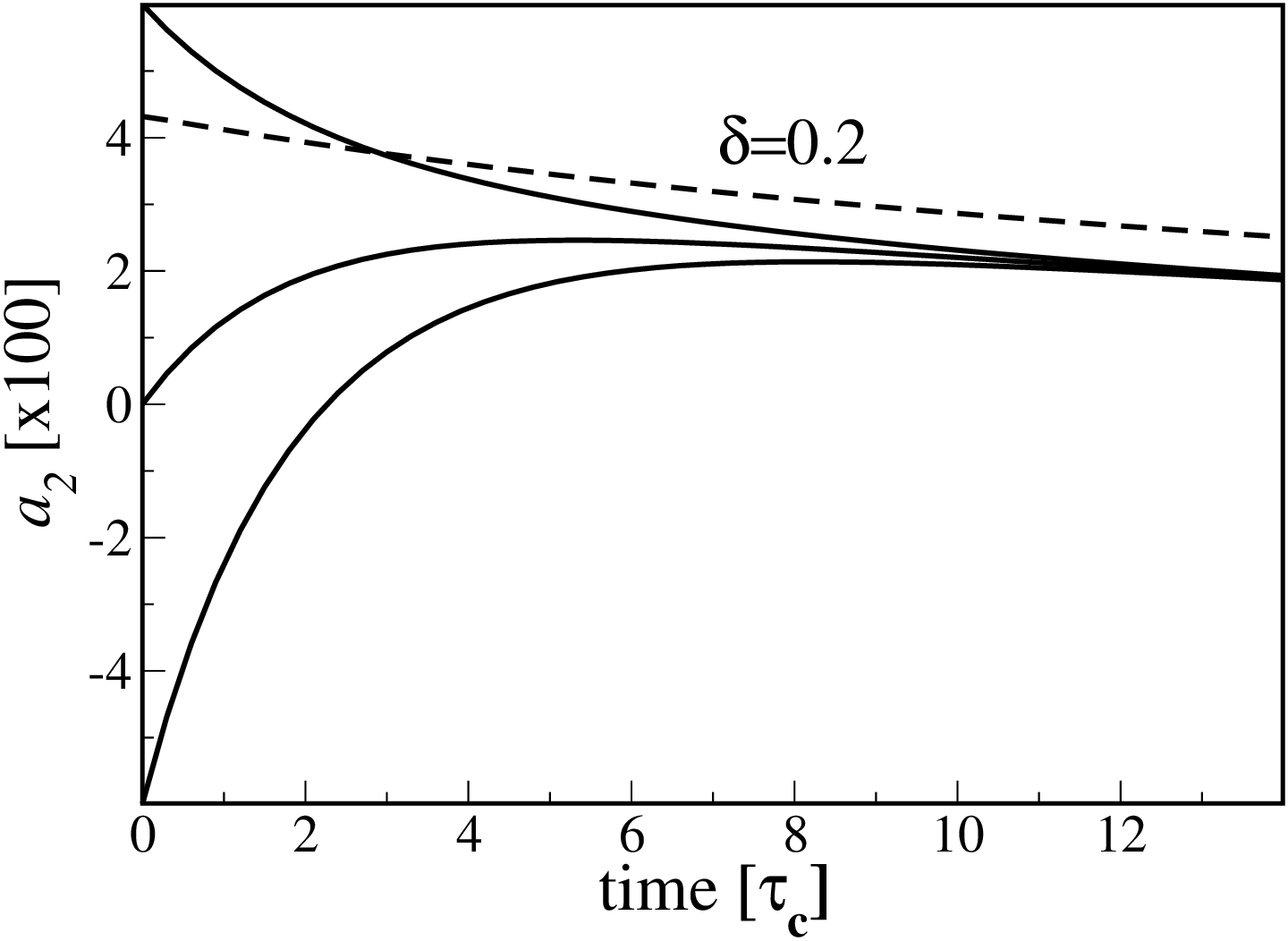}}
  \caption{The same as in Fig. \ref{fig:a2shortwLi}, but for different initial values of $a_2$.  The fast relaxation takes place during the first 5-10 collisions, independently on $a_2(0)$.}
  \label{fig:a2init}
\end{figure}

\begin{figure}[htbp]
   \centerline{\includegraphics[width=8cm,clip=]{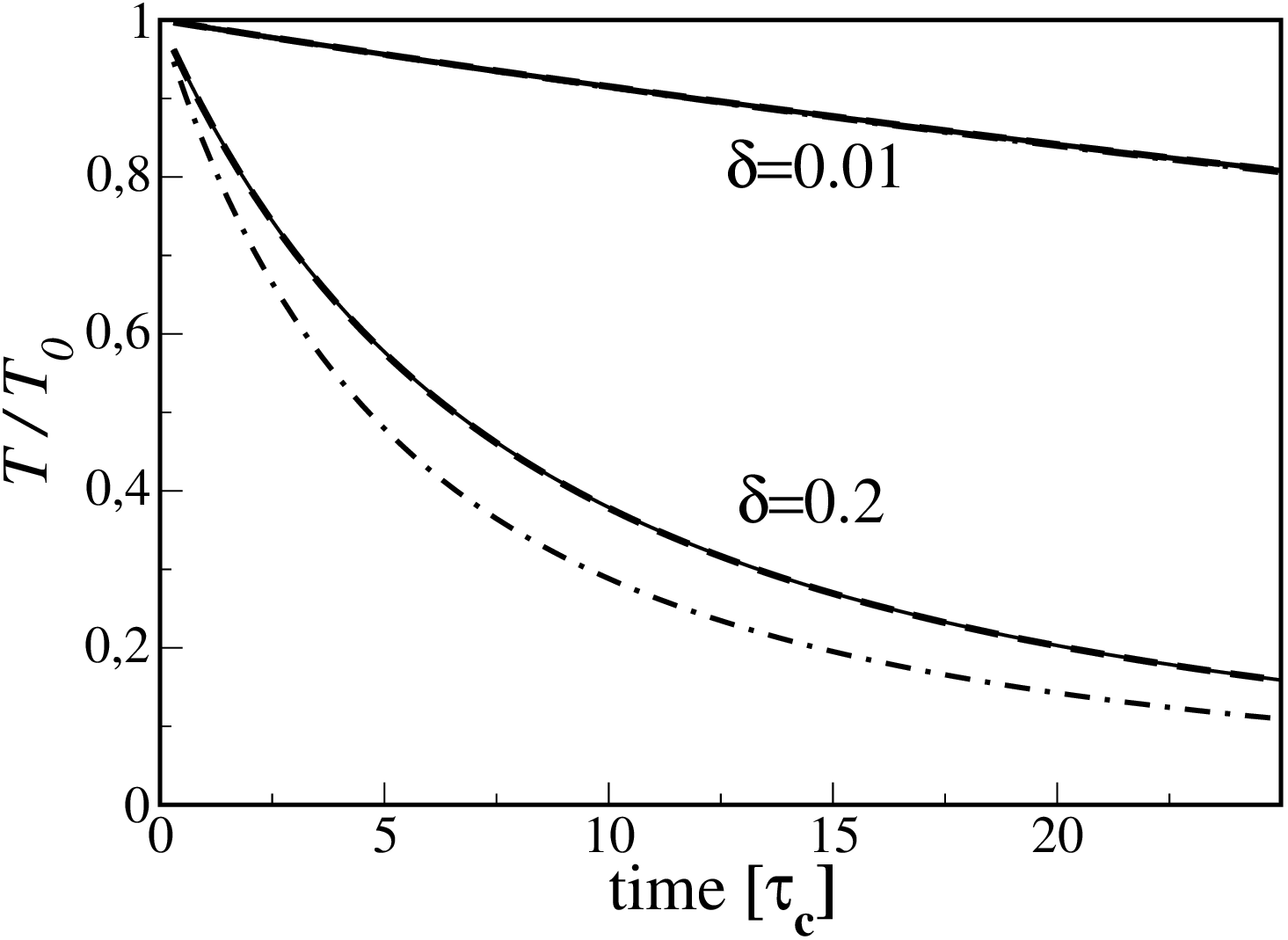}}
\vspace*{0.2cm}
   \centerline{\includegraphics[width=8cm,clip=]{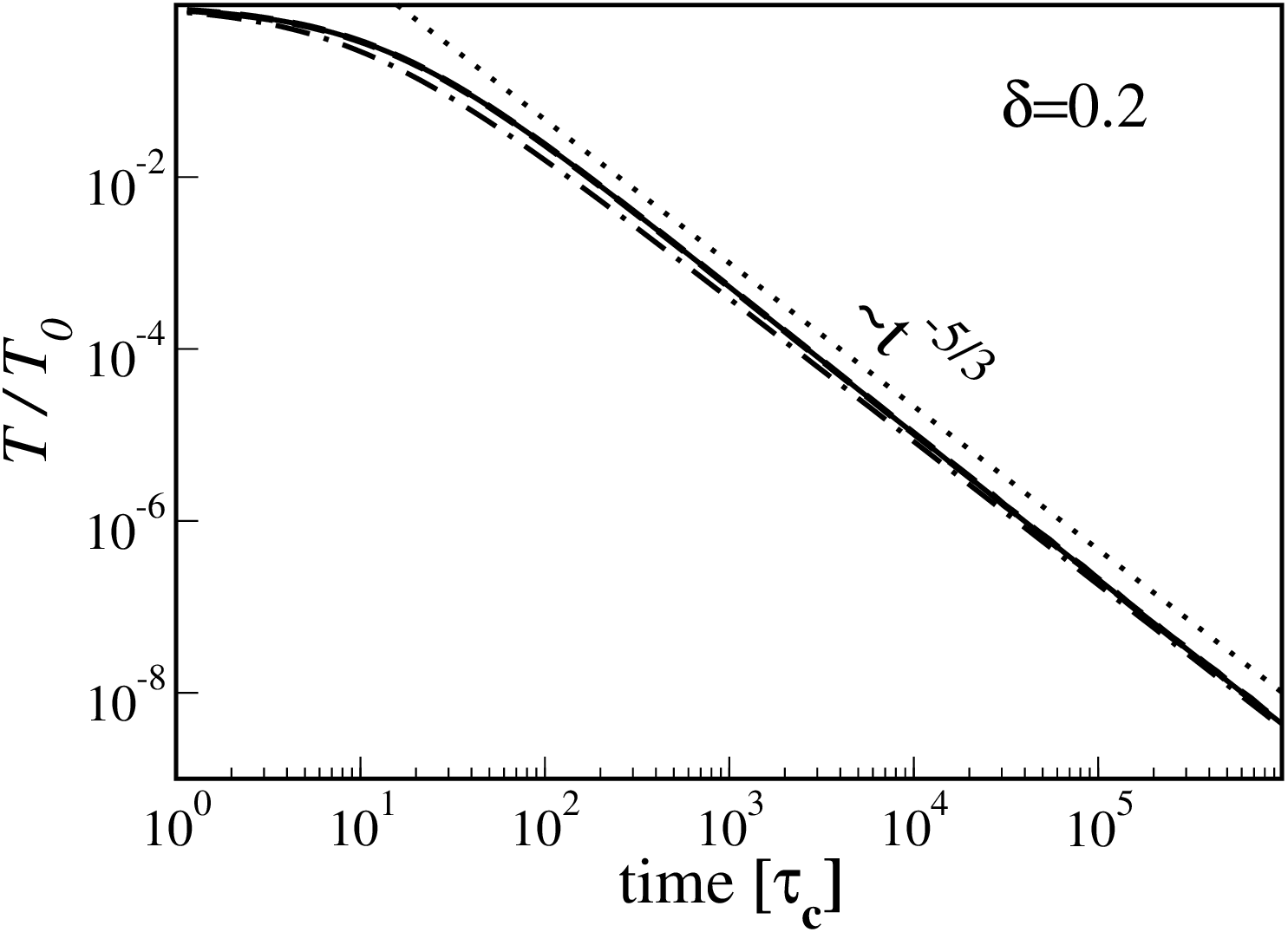}}
  \caption{Evolution of temperature in the homogeneous cooling state on the short time scale (top) and for longer time (bottom). The notations are the same as in Fig. \ref{fig:a2shortwLi}. The dotted line shows the asymptotic power-law  at $t \to \infty$.}
  \label{fig:Taiab}
\end{figure}

For the hydrodynamic description of granular gases we assume that there exist well separated time and length scales. The short time and length scales are given by the mean collision time and the mean free path, and the long time and length scales are characterized by the evolution of the hydrodynamic fields (to be  defined in the next section) and their spatial inhomogeneities.
The hydrodynamic approach corresponds to the coarse-grained description of the system, where all processes which take place on the short time and length scales are neglected. Therefore the first stage of the relaxation of the velocity distribution function does not affect the hydrodynamic description and only the second stage of its evolution on the time scale $\tau_0 \gg \tau_c(0)$ is to be taken into account.  
To this end we apply an adiabatic approximation: We omit the term $da_2/dt$ in the left-hand side of Eq. \eqref{eqa2} which describes the fast relaxation and assume that $a_2$ is determined by the current values of $\mu_2$ and $\mu_4$ due to the present temperature. Hence, in the adiabatic approximation $a_2$ is determined by
\begin{equation}
\label{eq:eqa2adiab}
5\mu_2 \left(1+a_2 \right)-\mu_4 =0 \, .
\end{equation}
Using \eqref{mu2A} for $\mu_2$, $\mu_4$ we find $a_2$ as an expansion in the small parameter $\delta^{\, \prime}$:
\begin{eqnarray}
\label{a2adiab}
a_2    &=& a_{21} \delta^{\, \prime}  + a_{22} \delta^{\, \prime \, 2}  + \dots \\
a_{21} &=& -\frac{3  \omega_0}{20 \sqrt{2 \pi}} \approx  -0.388 \nonumber \\
a_{22} &=& \frac{12063}{640000} \frac{\omega_0^2}{\pi} + \frac{27}{40} \frac{\omega_1}{\sqrt{2 \pi}} \approx 2.752 \, .\nonumber
\end{eqnarray}
Figures \ref{fig:a2shortwLi}  and \ref{fig:a2longwLi} show $a_2$ in adiabatic approximation due to Eq. \eqref{a2adiab} together with the numerical solution of Eqs. (\ref{dTdt},\ref{eqa2}) and the result of the linear theory, Eqs.(\ref{a21gensolLi},\ref{eq:defwt}). 
The adiabatic approximation is rather accurate for small dissipation, $\delta <0.05$, after the initial relaxation (5-10 collisions per particle)  has passed. Even for the larger dissipation ($\delta =0.2$) it is in agreement with the numerical result. This value of the dissipative parameter corresponds to the initial coefficient of restitution $\varepsilon \approx 0.75$ for the thermal velocity. The adiabatic approximation becomes more and more accurate as the system evolves. Contrary, the linear theory fails with increasing $\delta$ and becomes qualitatively incorrect for larger $\delta$. In Figs. \ref{fig:a2shortwLi} and \ref{fig:a2longwLi} the evolution of the system has started from a Maxwell distribution, i.e., $a_2(0)=0$.
Fig. \ref{fig:a2init} shows the relaxation of $a_2$ from certain initial conditions $a_2(0) \neq 0$ which correspond to non-Maxwellian distribution functions. The relaxation occurs during the first 5-10 collisions per particle. Then the adiabatic approximation becomes valid even for intermediate values of the dissipative parameter $\delta$. The adiabatic approximation yields very accurate results for the evolution of temperature. It coincides almost perfectly with the numerical solution for the short and for the long time scales and for all values of the dissipative parameter $0 < \delta \le 0.2$ (see Fig. \ref{fig:Taiab}). Thus we conclude that the adiabatic approximation may be applied for the hydrodynamic description of granular gases. 
By means of Eq. \eqref{a2adiab} for the second Sonine coefficient we can determine the moments $\mu_2$, $\mu_4$, and the cooling coefficient $\zeta$: 
\begin{equation}
\label{zetavisco}
\zeta=\frac23\mu_2 B  
= \frac23 \sqrt{\frac{2T}{m}} n \sigma^2 g_2(\sigma)  \left( \omega_0 \delta^{\, \prime} - \omega_2 \delta^{\, \prime \, 2}+\dots \right)\, ,
\end{equation}
where 
\begin{equation}
\omega_2 \equiv \omega_1+ \frac{9}{500}\omega_0^2 \sqrt{\frac{2}{\pi}}\,.  
\end{equation}
 
Below we will need the derivatives of $a_2$ and $\zeta$ with respect to temperature and density. Using 
\begin{equation}
\label{eq:derdelprime}
T \frac{\partial \delta^{\, \prime}}{\partial T}   = \frac{\delta^{\, \prime}}{10}  
\end{equation}
according to the definition of of $\delta^{\, \prime}$, Eq. \eqref{deltaprime}, we obtain
\begin{eqnarray}
\label{eq:Tda2dT}
T \frac{\partial a_2}{\partial T} &=& \frac{1}{10}a_{21} \delta^{\prime} + \frac{1}{5} a_{22} \delta^{\prime \, 2} \\ 
\label{eq:TdzetdT}
T \frac{\partial \zeta}{\partial T} &=& \frac23 B \left( \frac35 \omega_0 \delta^{\, \prime} 
                     - \frac{7}{10} \omega_2 \delta^{ \, \prime \, 2} +  \dots \right)\\
\frac{\partial \zeta}{\partial n} &=& \frac{1}{n} \zeta^{(0)} \, .
\label{zetdevisco}  
\end{eqnarray}

\section{Hydrodynamic equations and transport coefficients} 
\label{sec:Hydro}
\subsection{Derivation of hydrodynamic equations}
For the derivation of hydrodynamic equations from the Boltzmann equation it is assumed that there exist well separated time and length scales. As already briefly discussed in the preceding Section, in granular gases there are (at least) two sets of scales: The microscopic scales are characterized by the mean collision time and the mean free path. The macroscopic scales are  given by the characteristic time of the evolution of the hydrodynamic fields and the size of their spatial inhomogeneities. Hence, scale separation means that the macroscopic fields vary very slowly in space and time if measured in the microscopic units. This assumption allows for a gradient expansion in space and time, i.e. the application of the Chapman-Enskog  method \cite{ChapmanCowling:1970}.

The application of the Chapman-Enskog approach to granular gases is more sophisticated than its application to molecular gases: For molecular gases the unperturbed solution (the basic solution) for the velocity distribution function is the Maxwell distribution while for granular gases the basic solution is the  time-dependent velocity distribution of the homogeneous cooling state. 

For gases of particles in the homogeneous cooling state, which collide with $\varepsilon=\mbox{const.}$, the velocity distribution function  depends on time only through the time-dependent characteristic velocity. The shape of the velocity distribution function, given by $\tilde{f}$, is fixed by $\varepsilon$ and does not depend on time. Hence the evolution of temperature describes the evolution of the velocity distribution function exhaustively. For granular gases of viscoelastic particles the shape of the velocity distribution function, characterized by the Sonine coefficients $a_k$, is time dependent due to the time dependence of these coefficients. In the previous section we exemplified this property for the first non-trivial coefficient $a_2(t)$, assuming that for small dissipation higher coefficients may be disregarded.
Therefore, the evolution of granular gases of viscoelastic particles is described by the time dependence of temperature, i.e., by the second moment of the velocity distribution function, and by the time dependence of its higher-order even moments which are characterized by the Sonine coefficients $a_2$, $a_3$, etc. [see Eq. \eqref{eq:SonineMoments}]. 
Hence, the hydrodynamic description of granular gases whose particles collide with $\varepsilon = \varepsilon(g)$ requires an extended set of hydrodynamic fields which includes the higher-order moments.  A regular approach to derive hydrodynamic equations for this extended set of fields is the Grad method \cite{Grad:1949}. 
Alternatively, for small dissipation an adiabatic approximation can be applied: It is assumed that the shape  of the velocity distribution function, albeit varying in time, follows adiabatically the current temperature. Correspondingly, the higher-order moments are determined  by the temperature too. With this approximation a closed set of hydrodynamic equations for  density $n(\vec{r},t)$, velocity $\vec{u}(\vec{r},t)$, and temperature $T(\vec{r},t)$ may be derived. These fields are defined respectively  by the zeroth, first, and second moments of the velocity distribution function:
\begin{eqnarray}
\label{defnuT} 
  \begin{split}
    n\left(\vec{r}, t\right)                       & =  \int d \vec{v} f\left(\vec{r},\vec{v},t\right)\\
    n\left(\vec{r}, t\right)\, \vec{u}\left(\vec{r}, t\right)   & =  \int d \vec{v}\, \vec{v} f\left(\vec{r},\vec{v},t\right) \, , \\
    \frac32 n\left(\vec{r}, t\right)\, T\left(\vec{r}, t\right) & =  \int d \vec{v}\, \frac12 m V^2 f\left(\vec{r},\vec{v},t\right) \,,
  \end{split}
\end{eqnarray}
where $\vec{V}\equiv \vec{v}-\vec{u}(\vec{r}, t)$. For small dissipation the velocity distribution is well approximated by the second Sonine coefficients $a_2(t)$, i.e., higher-order coefficients are neglected. In adiabatic approximation $a_2$ is determined by temperature according to \eqref{a2adiab}.  

For simplicity of the notation we consider a dilute granular gas and approximate the Enskog factor $g_2(\sigma) \approx 1$. Multiplying the Boltzmann equation for an inhomogeneous gas
\begin{equation}
\label{2BoltzLor}
\left( \frac{\partial}{\partial t} + \vec{v}_1 \cdot
\vec{\nabla} 
\right)f(\vec{r}, \vec{v}_1, t)
=I\left(f,f \right) \, ,
\end{equation}
correspondingly by $v_1^0$, $\vec{v}_1$, $mv_1^2/2$, and integrating over $d \vec{v}_1$ we obtain the hydrodynamic equations (see e.g. \cite{ChapmanCowling:1970})
\begin{eqnarray}
\label{hydron}
\frac{\partial n}{\partial t} + \vec{\nabla} \cdot \left( n
\vec{u}\, \right)=0  \, , \\
\label{hydrou}
\frac{\partial \vec{u}}{\partial t} +\vec{u} \cdot
\vec{\nabla} \vec{u} +(nm)^{-1} 
\vec{\nabla}\cdot \hat{P}=0 \, , \\
\label{hydroT}
\frac{\partial T}{\partial t} +\vec{u} \cdot
\vec{\nabla} T 
+ \frac{2}{3n} \left( \hat{P}: \vec{\nabla} \vec{u} + 
\vec{\nabla} \cdot \vec{q} \right) + \zeta T =0   \,.
\end{eqnarray}
The cooling coefficient $\zeta $ in the sink term $\zeta T$ may be written as 
\begin{multline}
\label{eq:zeta_def}
\zeta\left(\vec{r}, t\right) = \frac{\sigma^2 m}{12nT} \int d\vec{v}_1 \int d\vec{v}_2 \int d\vec{e}\, \Theta\left(-\vec{v}_{12} \cdot \vec{e}\,\right) 
\left|\vec{v}_{12} \cdot \vec{e}\,\right|\\ 
 \times  f\left(\vec{r},\vec{v}_1,t\right) f\left(\vec{r},\vec{v}_2,t\right) 
  \left(\vec{v}_{12} \cdot \vec{e}\, \right)^2
   \left(1-\varepsilon^2\right)  \, .  
\end{multline}
The pressure tensor $\hat{P}$ and the heat flux $\vec{q}$ are defined by
\begin{eqnarray}
\label{eq:P_and_q}
P_{ij} \left(\vec{r},t\right) & = & \int D_{ij} \left(\vec{V}\right) f\left(\vec{r},\vec{v},t\right) d\vec{v} + p  \delta_{ij}\\
\vec{q}\left(\vec{r},t\right) & = & \int \vec{S} \left(\vec{V}\right)   f\left(\vec{r},\vec{v},t\right) d\vec{v}\,,
\end{eqnarray}
where $p=nT$ is the hydrostatic pressure. The velocity tensor $D_{ij}$ and the vector $\vec{S}$ read 
\begin{eqnarray}
\label{eq:def_Dij}
D_{ij} (\vec{V}) &\equiv& m \left( V_iV_j -\frac13 \delta_{ij} V^2 \right)\\
\label{1qdef}
\vec{S} \left(\vec{V}\right) &\equiv& \left( \frac{mV^2}{2} - \frac52 T \right)\vec{V} \, .
\end{eqnarray} 
The structure of the hydrodynamic equations, except for the cooling term $\zeta T$, coincides with those for molecular gases. 

\subsection{Chapman-Enskog approach}
The system  (\ref{hydron},\ref{hydrou},\ref{hydroT}) is closed by expressing the pressure tensor and the heat flux in terms of the hydrodynamic fields and the fields gradients. To this end we apply the Chapman-Enskog approach \cite{ChapmanCowling:1970}.
This method is based on two important assumptions: 
The evolution of the distribution function is completely determined by the evolution of its first few moments, i.e., it depends on space and time only through the hydrodynamic fields:
\begin{equation}
\label{eq:evolVDF}
f\left(\vec{r},\vec{v},t\right) = f\left[ \vec{v}, n\left(\vec{r},t\right), \vec{u}\left(\vec{r},t\right), T\left(\vec{r},t\right) \right] \, .
\end{equation} 
As the second precondition it is assumed  that the gas is only slightly inhomogeneous on the microscopic length scale which allows for a gradient expansion of the velocity distribution function:
\begin{equation}
\label{fserlamb}
f=f^{(0)}+\lambda f^{(1)} +\lambda^2 f^{(2)} + \dots \, , 
\end{equation} 
where each power $k$ of the formal parameter $\lambda$ corresponds to the order $k$ in the spatial gradient. Thus, $f^{(0)}$ refers to the homogeneous cooling state, $f^{(1)}$ corresponds to the linear approximation with respect to the fields gradients, $f^{(2)}$ is the solution with respect to quadratic terms in the field gradients, etc. With these assumptions the Boltzmann equation may be solved iteratively for each order in $\lambda$, together with the hydrodynamic equations for the moments of the velocity distribution function. The solution in zeroth order in $\lambda$ yields the velocity distribution function for the homogeneous cooling state $f^{(0)}$ and the corresponding evolution of temperature. The function $f^{(0)}$ is then used to compute $\hat{P}$ and $\vec{q}$, yielding $\hat{P}^{(0)}=\delta_{ij}p$, $\vec{q}^{\,(0)}=0$ and the hydrodynamic equations for the ideal fluid. These first-order equations contain only linear gradient terms. Then  $f^{(1)}$ may be found employing the first-order hydrodynamic equations and the distribution function $f^{(0)}$. The obtained $f^{(1)}$  as well as the corresponding expressions for $\hat{P}^{(1)}$ and $\vec{q}^{\,(1)}$ are linear in the field gradients:
\begin{eqnarray}
\label{Petaqkappa}
\begin{split}
P_{ij}=&p \delta_{ij} - 
\eta \left( \nabla_i u_j +\nabla_j u_i - 
\frac23 \, \delta_{ij} \vec{\nabla} \cdot \vec{u} \right)\\
\vec{q}= &-\kappa \vec{\nabla} T - \mu \vec{\nabla}
n \, . 
\end{split}
\end{eqnarray}
The transport coefficients $\eta$, $\kappa$ and $\mu$ in these equations are expressed in terms of $f^{(1)}$. Hence, within the Chapman-Enskog approach for each order in the gradient expansion a closed set of equations may be derived. Keeping only first-order field gradients for the distribution function and respectively  for the pressure tensor and for the heat flux as in Eq. \eqref{Petaqkappa}, the Navier-Stokes hydrodynamics is obtained. Keeping next-order gradient terms corresponds to the Burnett or super-Burnett description. We will restrict ourselves to the Navier-Stokes level \footnote{It has been shown that some processes in granular gases require the Burnett level to derive a consistent set of equations \cite{GoldhirschGG:2000,SelaGoldhirsch:1998}.} and skip for simplicity of the notation the superscript  ``(1)'' for  $\hat{P}$ and $\vec{q}$ in the above equations. 

The Chapman-Enskog scheme also assumes a hierarchy of time scales and respectively a hierarchy of time derivatives:  
\begin{equation}
\label{tdervialamb}
\frac{\partial}{\partial t} =  \frac{\partial^{(0)}}{\partial t} + \lambda \frac{\partial^{(1)}}{\partial t} + \lambda^2 \frac{\partial^{(2)}}{\partial t}+ \dots  \, , 
\end{equation}
where  each order $k$ in the time derivative, $ \partial^{(k)}/\partial t  \equiv \partial_t^{(k)}$ corresponds to the related order in the space gradient. 
Consequently, the higher the order in the space gradient, the slower is the according time variation. Using the formal expansion parameter $\lambda$ we can write the Boltzmann equation, 
\label{page:d_dt_01k}
\begin{multline}
\label{Boltzexpan}
\left( \frac{\partial^{(0)}}{\partial t} + \lambda \frac{\partial^{(1)}}{\partial t} + \dots \lambda \vec{v}_1 \cdot \vec{\nabla} \right) \left( f^{(0)}+\lambda f^{(1)} + \dots \right) \\
=I\left[ \left( f^{(0)}+\lambda f^{(1)}  + \dots \right), \left( f^{(0)}+\lambda f^{(1)}  + \dots \right) \right]  
\end{multline}
and collect terms of the same order in $\lambda$. The equation in zeroth order
\begin{equation}
\label{BEzero}
\frac{\partial^{(0)}}{\partial t} f^{(0)} = I\left(f^{(0)}, f^{(0)} \right) 
\end{equation} 
coincides with Eq. \eqref{collint1} for the homogeneous cooling state. According to the Chapman-Enskog scheme we obtain the velocity distribution function in zeroth order:
\begin{equation}
\label{fzero}
f^{(0)}\left(\vec{v}, \vec{r}, t\right) =\frac{n\left(\vec{r},t\right)}{v_T^3\left(\vec{r},t\right)} \left[1+a_2 S_2\left(c^2\right) \right] \phi(c) \, ,  
\end{equation} 
where $\vec{c}=\left(\vec{v}-\vec{u}\right)/v_T=\vec{V}/v_T$. The corresponding hydrodynamic equations in this order read
\begin{equation}
\label{hydrozero}
\frac{\partial^{(0)}}{\partial t} n=0, \qquad 
\frac{\partial^{(0)}}{\partial t} \vec{u} =0, \qquad 
\frac{\partial^{(0)}}{\partial t} T = -\zeta^{(0)} T \, , 
\end{equation} 
where $\zeta^{(0)}$ is to be calculated using Eq. \eqref{eq:zeta_def} with  $f=f^{(0)}$. In this way we reproduce the previous result \eqref{zetavisco}. 

Collecting terms of the order ${\cal O} ( \lambda ) $ we obtain, 
\begin{equation}
\label{BEfirst}
\frac{\partial^{(0)} f^{(1)}}{\partial t} +\left( \frac{\partial^{(1)}}{\partial t} + \vec{v}_1 \cdot \vec{\nabla} \right) f^{(0)} + 
J^{(1)}\left(f^{(0)},f^{(1)}\right) =0 \, , 
\end{equation} 
where we introduce 
\begin{equation}
\label{eq:defJofI}
-J^{(1)}\left(f^{(0)},f^{(1)}\right) \equiv I\left( f^{(0)}, f^{(1)} \right) +I\left(f^{(1)}, f^{(0)} \right) \, .
\end{equation} 
The corresponding first-order hydrodynamic equations read
\begin{eqnarray}
\label{hydrofirst}
\begin{split}
\frac{\partial^{(1)}n}{\partial t}   =&  - \vec{\nabla} \left( n \vec{u} \right)\\
\frac{\partial^{(1)} \vec{u}}{\partial t} =& - \vec{u} \cdot  \vec{\nabla} \vec{u} - \frac{1}{nm} \vec{\nabla} p \\
\frac{\partial^{(1)} T}{\partial t}   =&  - \vec{u} \cdot \vec{\nabla}   T  - \frac23  T \, \vec{\nabla}  \cdot \vec{u}  -\zeta^{(1)} T  \,.
\end{split}
\end{eqnarray}
The term $\zeta^{(1)}$ is found by substituting $f=f^{(0)}+\lambda f^{(1)}$ into  Eq. \eqref{eq:zeta_def} and collecting terms of the order ${\cal O} ( \lambda )$. These terms contain the factor $f^{(0)} \left(\vec{v},\vec{r},t\right)f^{(1)} \left(\vec{v},\vec{r},t\right)$ in the integrand. They vanish  upon integration according to the different symmetry of the functions  $f^{(0)}$ and $f^{(1)}$, thus, $\zeta^{(1)}=0$  \footnote{This may be directly checked using Eq. (\ref{f1form}) for $f^{(1)}$}. 

Since the distribution function $f^{(0)}$ is known, we can evaluate those terms in \eqref{BEfirst} which depend only on $f^{(0)}$: 
\begin{multline}
\label{df0dtfirst}
\left( \frac{\partial^{(1)}}{\partial t} + \vec{v}_1 \cdot \vec{\nabla} \right) f^{(0)} = \frac{\partial f^{(0)} }{\partial n} \left( \frac{\partial^{(1)} n}{\partial t} + \vec{v}_1 \cdot \vec{\nabla} n \right)\\
+\frac{\partial f^{(0)} }{\partial \vec{u}} \cdot \left( \frac{\partial^{(1)} \vec{u}}{\partial t} + \vec{v}_1 \cdot \vec{\nabla} \vec{u} \right) \\
+ \frac{\partial f^{(0)}}{\partial T}  \left( \frac{\partial^{(1)}}{\partial t} T + \vec{v}_1 \cdot \vec{\nabla} T \right)  \, . 
\end{multline}
With $\vec{v}_1=\vec{V}+\vec{u}$, the time derivatives of $n$, $\vec{u}$ and $T$ given in \eqref{hydrofirst}, and the relations 
\begin{equation}
\label{f0nuTder}
\frac{\partial f^{(0)} }{\partial n} = \frac{1}{n} f^{(0)}, \qquad
\frac{\partial f^{(0)} }{\partial \vec{u} } = -\frac{\partial f^{(0)} }{\partial \vec{V} } \, ,
\end{equation}
which follow from Eq. \eqref{fzero}, we recast Eq. \eqref{BEfirst} for $f^{(1)}$ into the form: 
\begin{multline}
\label{11BEfirst}
\frac{\partial^{(0)} f^{(1)}}{\partial t} + J^{(1)}\left( f^{(0)},f^{(1)} \right) 
= f^{(0)} \left( \vec{\nabla} \cdot \vec{u} - \vec{V} \cdot \vec{\nabla} \log n \right) \\ 
+\frac{\partial f^{(0)}}{\partial T} \left( \frac{2}{3} \, T \vec{\nabla} \cdot \vec{u} - \vec{V} \cdot \vec{\nabla} T\right) \\ 
+\frac{\partial f^{(0)} }{\partial V_i} \left( \left(\vec{V} \cdot \vec{\nabla}\right)u_i 
       -\frac{1}{nm}\nabla_i \, p \right) \, , 
\end{multline}
where we take into account $\zeta^{(1)}=0$.

The right-hand side is known since it contains only $f^{(0)}$. It is convenient, however to rewrite it in a form  which shows explicitly the dependences on the fields gradients. Employing
\begin{equation}
\frac{1}{nm}  \nabla_i p = \frac{T}{m} \nabla_i \log T +  \frac{T}{m} \nabla_i \log n \, , 
\end{equation} 
the right-hand side of \eqref{11BEfirst} yields
\begin{multline}
\label{3BEfirst}
\frac{\partial^{(0)} f^{(1)}}{\partial t} + J^{(1)}\left(f^{(0)},f^{(1)}\right) \\
=\vec{A} \cdot \vec{\nabla} \log T + \vec{B} \cdot \vec{\nabla} \log n + C_{ij} \nabla_j u_i \, ,  
\end{multline} 
with
\begin{eqnarray}
\label{coefABCvisco1}
\vec{A}(\vec{V}) & =& -\vec{V} T \frac{\partial f^{(0)}}{\partial T} -\frac{T}{m} \frac{\partial f^{(0)}}{\partial \vec{V}}  \\
\label{coefABCvisco2}
\vec{B}(\vec{V}) & =& -\vec{V} f^{(0)} -  \frac{T}{m} \frac{\partial }{\partial \vec{V}} f^{(0)} \\
\label{coefABCvisco3}
C_{ij}(\vec{V}) & =&  \frac{\partial  }{\partial V_i} \left( V_j f^{(0)} \right) +\frac{2T}{3} \delta_{ij} \frac{\partial f^{(0)}}{\partial T} \,. 
\end{eqnarray}
We will calculate these terms below. From the form of the right-hand side of Eq. \eqref{3BEfirst} we expect the form of its solution  
\begin{equation}
\label{f1form}
f^{(1)} = \vec{\alpha} \cdot \vec{\nabla} \log T + \vec{\beta} \cdot \vec{\nabla} \log n + \gamma_{ij} \nabla_j u_i \,,
\end{equation}
which is the most general form of a scalar function, which depends linearly on the vectorial  gradients $ \vec{\nabla} T$, $\vec{\nabla} n$ and on the tensorial gradients $\nabla_j u_i$. The coefficients $\vec{\alpha}$, $\vec{\beta}$ and $\gamma_{ij}$ are functions of $\vec{V}$ and of the hydrodynamic fields $n$, $\vec{u}$  and $T$. 

We derive now equations for the coefficients $\vec{\alpha}$, $\vec{\beta}$ and $\gamma_{ij}$ by substituting $f^{(1)}$ as given by Eq. \eqref{f1form} into the first-order equation \eqref{3BEfirst} and equating the coefficients of the corresponding gradients. To this end we need $\partial_t^{(0)} f^{(1)}$ and, therefore, the time derivatives of the coefficients $\vec{\alpha}$, $\vec{\beta}$, and $\gamma_{ij}$:
\begin{multline}
\label{dalphadt}
\frac{\partial^{(0)}\vec{\alpha}}{\partial t} =\frac{\partial \vec{\alpha} }{\partial T}  \frac{\partial^{(0)} T}{\partial t} + \frac{\partial \vec{\alpha} }{\partial n } \frac{\partial^{(0)} n}{\partial t} + \frac{\partial \vec{\alpha} }{\partial u_i} \frac{\partial^{(0)} T}{\partial t}\\
= -\zeta^{(0)} T \frac{\partial \vec{\alpha}}{\partial T}\, ,
\end{multline}
where we use Eqs. \eqref{hydrozero} in zeroth order. Similarly we obtain 
\begin{equation}
\label{dbetgamdt}
  \frac{\partial^{(0)}\vec{\beta}}{\partial t}  = -\zeta^{(0)} T \frac{\partial \vec{\beta}}{\partial T}\,, 
\qquad
\frac{\partial^{(0)}\gamma_{ij}}{\partial t}  =  -\zeta^{(0)} T \frac{\partial \gamma_{ij}}{\partial T}
\end{equation}
and, respectively, the time-derivatives of the gradients
\begin{equation}
\label{dgradTnudt}
\begin{split}
\frac{\partial^{(0)}}{\partial t} \vec{\nabla} \log n  = & 0\\
\frac{\partial^{(0)}}{\partial t} \nabla_j u_i = & 0 \\
\frac{\partial^{(0)}}{\partial t} \vec{\nabla} \log T  = &  - \vec{\nabla} \zeta^{(0)}\\
  =&-\left( \frac{\partial \zeta^{(0)} }{\partial n} \right)\vec{\nabla} n  -\left( \frac{\partial \zeta^{(0)} }{\partial T} \right)\vec{\nabla} T\,.
\end{split}
\end{equation}
The derivatives of $\zeta^{(0)}$ are given by Eqs. (\ref{eq:TdzetdT},\ref{zetdevisco}). From \eqref{dalphadt}, \eqref{dbetgamdt}, and \eqref{dgradTnudt} we obtain:
\begin{multline}
\label{df1dt0}
\frac{\partial^{(0)} f^{(1)}}{\partial t} = -\left(T \frac{\partial \zeta^{(0)}}{\partial T} \vec{\alpha} \right)  \cdot \vec{\nabla} \log T \\
- \left( \zeta^{(0)} T \frac{\partial \vec{\beta}}{\partial T} + \zeta^{(0)} \vec{\alpha} \right) \cdot \vec{\nabla} \log n -  \zeta^{(0)} T \frac{\partial \gamma_{ij}}{\partial T} \nabla_j u_i\,.
\end{multline}
where we use Eq.(\ref{zetdevisco}).  If we insert $\partial_t^{(0)} f^{(1)}$ into Eq. \eqref{3BEfirst} and equate the coefficients of the gradients we arrive at a set of equations for the coefficients  $\vec{\alpha}$, $\vec{\beta}$ and $\gamma_{ij}$: 
\begin{eqnarray}
\label{eqsalbega1}
-T \frac{\partial \zeta^{(0)}}{\partial T}  \vec{\alpha} +J^{(1)}\left(f^{(0)}, \vec{\alpha} \right)   & = \vec{A} \\
\label{eqsalbega2}
-\zeta^{(0)} T \frac{\partial \vec{\beta}}{\partial T} - \zeta^{(0)} \vec{\alpha} +J^{(1)}\left(f^{(0)}, \vec{\beta}\, \right)  & =  \vec{B} \\
\label{eqsalbega3}
-\zeta^{(0)} T \frac{\partial \gamma_{ij}}{\partial T} +J^{(1)}\left(f^{(0)},  \gamma_{ij} \right)            & =  C_{ij}   \, 
\end{eqnarray}
where $J^{(1)}$ is defined by Eq. \eqref{eq:defJofI}.

\subsection{Kinetic coefficients in terms of the velocity distribution function}
From the definition of the pressure tensor Eq. \eqref{eq:P_and_q} and its expression in terms of the field gradients Eq. \eqref{Petaqkappa} follows
\begin{multline}
\label{Dijeta}
\int D_{ij} 
\left(f^{(0)} + \vec{\alpha} \cdot \vec{\nabla} \log T + \vec{\beta} \cdot \vec{\nabla} \log n + 
\gamma_{kl} \nabla _l u_k \right) d \vec{V} \\
 = - \eta \left( \nabla_i u_j + \nabla_j u_i -\frac23 \delta_{ij} \vec{\nabla} \cdot \vec{u} \right) \, ,
\end{multline}
where the tensor $D_{ij}\left( \vec{V} \right) $ has been defined above. The integrals
\begin{equation}
\label{Dalbeteq0}
\int D_{ij} f^{(0)}d \vec{V}
=\int D_{ij} \vec{\alpha}\, d \vec{V} = \int D_{ij}  \vec{\beta}\, d \vec{V} =0
\end{equation}
vanish since $D_{ij}$ is a traceless tensor and $f^{(0)}$ depends isotropically on $\vec{V}$. Moreover, as will be shown below, the vectors $\vec{\alpha}$ and  $\vec{\beta}$ are directed along $\vec{V}$, hence,  the respective integrands are odd functions of $\vec{V}$. Therefore, only the term with the factor $\gamma_{kl} \nabla _l u_k$ in the left-hand side of Eq. \eqref{Dijeta} is non-trivial. Equating the coefficients of the gradient factor $\nabla _l u_k$ we obtain
\begin{equation}
\label{Dgameta}
\int D_{ij}  \gamma_{kl}  d \vec{V} = -\eta \left( \delta_{li}\delta_{kj}+\delta_{lj}\delta_{ki} -\frac23 \delta_{ij}\delta_{kl} \right)\,.
\end{equation}
For $k=j$, $l=i$ the last equation turns into 
\begin{equation}
\label{1Dgameta}
\int D_{ij} \gamma_{ji}  d \vec{V} = -\eta \left( \delta_{ii}\delta_{jj}+\delta_{ij}\delta_{ij} -\frac23 \delta_{ij}\delta_{ij} \right) =-10 \eta \, , 
\end{equation}
($\delta_{ii}\delta_{jj} =9$ and $\delta_{ij}\delta_{ij} =3$ according to the summation convention) and yields the coefficients of viscosity
\begin{equation}
\label{2Dgameta}
\eta=-\frac{1}{10} \int D_{ij}\left( \vec{V} \right) \gamma_{ji} \left( \vec{V} \right) d \vec{V}\,.
\end{equation}
Using Eq. \eqref{eq:P_and_q} for the heat flux and its corresponding expression in terms of the field gradients, Eq. \eqref{Petaqkappa}, we can perform a completely analogous calculation and arrive at the kinetic coefficients $\kappa$ and $\mu$:
\begin{eqnarray}
\label{kapSmuSfin}
\kappa & = & - \frac{1}{3T} \int d \vec{V} \vec{S}\left( \vec{V} \right) \cdot \vec{\alpha} \left( \vec{V} \right) \\
\mu  &   = & - \frac{1}{3n} \int d \vec{V} \vec{S}\left( \vec{V} \right) \cdot \vec{\beta}\left( \vec{V} \right) \,,
\end{eqnarray}
with $\vec{S}$ defined by Eq. \eqref{1qdef}.
The coefficient of thermal conductivity $\kappa$ has the standard interpretation, while the other coefficient $\mu$ does not have an analog for molecular gases. 

\section{Coefficient of viscosity}
The viscosity coefficient is related to the coefficient $\gamma_{ij}$ [see Eq.  \eqref{2Dgameta}] which, in its turn, is the solution of Eq. \eqref{eqsalbega3} with the coefficient $C_{ij}$ in the right-hand side. Let us first find an explicit expression for $C_{ij}$.

According to Eq. \eqref{fzero} the velocity distribution function depends on temperature through the thermal velocity $v_T$ and additionally through the second Sonine coefficient $a_2$. Hence the temperature derivative of $f^{(0)}$ reads
\begin{equation}
\label{eq:df0dtvisco}
\frac{\partial f^{(0)}}{\partial T} =\frac{1}{2T} \frac{\partial }{\partial \vec{V}} \cdot \vec{V} f^{(0)} +f_M (V) S_2\left(c^2\right) \frac{\partial a_2}{\partial T} \, , 
\end{equation}
with $f_M (V)$ being the Maxwell distribution
\begin{equation}
\label{eq:Maxwdistr}
  f_M=\frac{n}{v_T^3} \phi(c)~~~~
\mbox{with}~~~~\phi(c)= \frac{1}{\pi^{3/2}} \exp \left( - {c^2} \right) \, . 
\end{equation}
Using the relation
\begin{equation}
\label{f0isotropic}
\frac{\partial f^{(0)} }{ \partial V_i} = \frac{V_i}{V} \frac{\partial f^{(0)} }{ \partial V} 
\end{equation}
and Eq. \eqref{eq:df0dtvisco} the coefficient $C_{ij}$ reads
\begin{multline}
\label{eq:Cij_visco}
C_{ij}\left(\vec{V}\right) = \left( V_iV_j -\frac13 \delta_{ij} V^2 \right) \frac{1}{V} \frac{\partial f^{(0)} }{ \partial V } \\
+ \frac23 \delta_{ij} S_2\left(c^2\right)f_M(V) \, T\frac{\partial a_2}{\partial T} \, . 
\end{multline}
With 
\begin{equation}
\label{eq:df0dV}
\frac{1}{V} \frac{\partial f^{(0)} }{ \partial V } 
             = -\frac{m}{T} \left[ 1 + a_2 S_2\left(c^2\right)  \right]\, f_{M}  
               + \frac{m}{T} \left( c^2 - \frac52 \right) a_2 \, f_{M}  \, 
\end{equation}
we obtain
\begin{multline}
\label{eq:Cij_eps_const}
C_{ij} = -\frac{1}{T} D_{ij}\left[ 1+ a_2 \left( S_2\left(c^2\right) + \frac52 -c^2 \right) \right] f_{M}(V) \\
+ \frac23 \delta_{ij} S_2\left(c^2\right)f_M(V) \, T\frac{\partial a_2}{\partial T} \, , 
\end{multline}
where $D_{ij}\left(\vec{V} \right)$ has been defined in Eq. \eqref{eq:def_Dij}. 

The expression for $C_{ij}$ determines the right-hand side of Eq. \eqref{eqsalbega3} for $\gamma_{ij}$ and hence it suggests the form for $\gamma_{ij}$. For small dissipation (when $a_2$ is small) and for small fields gradients we keep only the leading terms  with respect to these variables. Therefore, we seek for $\gamma_{ij}$ in the form
\begin{equation}
\label{1gamfirst}
\gamma_{ij}\left(\vec{V}\right) = \frac{\gamma_0 }{T} D_{ij}\left(\vec{V}\right)f_{M}(V) \, ,  
\end{equation}
where $\gamma_0$ is  a velocity-independent coefficient, i.e., we neglect the dependence of $\gamma_{ij}$ on $a_2$ \footnote{In a more general approach $\gamma_{ij}$ is represented as an expansion in orthogonal polynomials, with \eqref{1gamfirst} being the first order term \cite{ChapmanCowling:1970}.}. 
The viscosity coefficient Eq. \eqref{2Dgameta} reads then
\begin{equation}
    \eta= -\frac{\gamma_0}{10} \, \frac{1}{T} \int d \vec{V}  D_{ij} D_{ij} f_M = -\gamma_0 nT \, ,  
\label{visforelast}
\end{equation}
where we take into account that 
\begin{equation}
D_{ij}D_{ji}=\frac23 m^2V^4  
\end{equation}
according the definition of $D_{ij}$, Eq. \eqref{eq:def_Dij}, with the summation convention. Moreover, we have used the fourth moment of the Maxwell distribution.  From Eq. \eqref{visforelast} follows
\begin{equation}
\gamma_0=-\frac{\eta}{nT}\,.   
\end{equation}

Multiplying  Eq. \eqref{eqsalbega3} by $D_{ij}(\vec{V}_1)$, integrating over $\vec{V}_1$ and using Eq. \eqref{2Dgameta}   yields
\begin{multline}
\label{eqforeta}
  - 10 \zeta^{(0)} T \frac{\partial \eta}{\partial T} 
  = -\int d \vec{V}_1 D_{ij}\left(\vec{V}_1 \right)  C_{ij}\left(\vec{V}_1 \right) \\
  + \int d \vec{V}_1 D_{ij}\left(\vec{V}_1 \right) J^{(1)}\left(f^{(0)},  \gamma_{ij} \right)
   \, . 
\end{multline} 
To evaluate the first  term in the right-hand side we use Eq. \eqref{coefABCvisco3} for $C_{ij}$, the relation
\begin{equation}
\frac{\partial  }{\partial V_{i}}D_{ij}=
\frac{\partial  }{\partial V_i}m \left( V_iV_j -\frac13 \delta_{ij} V^2 \right)=
mV_j \left( 1 + \frac13 \delta_{ij} \right) \,,
\end{equation} 
the definition of temperature, Eq. \eqref{defnuT}, and notice that $D_{ij}\delta_{ij}=0$ since 
$D_{ij}$ is a traceless tensor [see Eq. \eqref{eq:def_Dij}]. 
Integration by parts then yields
\begin{eqnarray}
\label{eq:Dij_Cij}
&& \int d\vec{V}_1 D_{ij}\, C_{ij} = \\
&& = \int d\vec{V}_1 D_{ij} \frac{\partial}{\partial V_{1i}} V_{1j} f^{(0)} + \frac{2T}{3} \int d\vec{V}_1 D_{ij} \delta_{ij} \frac{\partial f^{(0)}}{\partial T} \nonumber \\
&& = \int d\vec{V}_1 f^{(0)} mV_{1j}V_{1j} \left( 1 + \frac13 \delta_{ij} \right) 
=\frac{10}{3} \int d\vec{V}_1 f^{(0)} mV_1^2  \nonumber \\ 
&& = 10nT  \,.  \nonumber 
\end{eqnarray}
For the second term in the right-hand side of \eqref{eqforeta} we use the definition of $J^{(1)}$, \eqref{eq:defJofI}, and obtain
\begin{multline}
\label{Icolforgam1}
\int d \vec{V}_1 D_{ij}  J^{(1)}\left(f^{(0)}, \gamma_{ij}\right) \\
= - \int  d \vec{V}_1 D_{ij}  I\left(f^{(0)},\gamma_{ij}\right) - \int d \vec{V}_1 D_{ij}  I\left(\gamma_{ij}, f^{(0)}\right) \,.
\end{multline}
We apply the property of the collision integral \cite{ChapmanCowling:1970} 
\begin{multline}
\label{Icolltrans1}
\int d \vec{V}_1 D_{ij}  I\left(f^{(0)},  \gamma_{ij}\right) 
=\int d \vec{V}_1 D_{ij}  I\left(\gamma_{ij}, f^{(0)}\right) \\
=\frac{\sigma^2}{2} \int d \vec{V}_1 \int d \vec{V}_2 f^{(0)}\left(\vec{V}_1\right) \gamma_{ij} \left(\vec{V}_2\right) \int d\vec{e} \, \Theta\left(-\vec{V}_{12} \cdot \vec{e}\,\right)\\
\times \left|\vec{V}_{12} \cdot \vec{e}\,\right| \Delta \left[D_{ij}\left(\vec{V}_1\right) + D_{ij}\left(\vec{V}_2\right) \right] \,, 
\end{multline}
where $\Delta \psi\left(\vec{v}_i \right) \equiv \psi\left(\vec{v}^{\,\prime}_i \right) - \psi\left(\vec{v}_i \right)$ denotes as previously the change of some quantity $\psi(\vec{v}_i)$ due to a collision. Equation \eqref{Icolforgam1} turns then into 
\begin{multline}
\label{Icolltrans3}
\int d \vec{V}_1 D_{ij} J^{(1)}\left(f^{(0)}, \gamma_{ij}\right) =-\sigma^2 \int d \vec{V}_1 \int d \vec{V}_2 f^{(0)} \left(\vec{V}_1\right)\\ 
\times \gamma_{ij}\left(\vec{V}_2\right) \int d\vec{e} \, \Theta\left(-\vec{V}_{12} \cdot \vec{e}\,\right) 
    \left|\vec{V}_{12} \cdot \vec{e}\,\right|\\ 
\Delta \left[D_{ij}\left(\vec{V}_1\right) + D_{ij}\left(\vec{V}_2\right) \right] \, .
\end{multline}
We write the factors in the last integral using the dimensionless velocities $\vec{V}_{1/2} =v_T \vec{c}_{1/2}$:
\begin{equation}
\begin{split}
\label{eq:dimlessDij}
D_{ij}\left(\vec{V}\right) = & mv_T^2 D_{ij}\left(\vec{c}\,\right)=mv_T^2 \left(c_{i}c_{j}-\frac13 \delta_{ij} c^2 \right) \\
\gamma_{ij}\left(\vec{V} \right) = & \frac{\gamma_0}{T} \left( \frac{n}{v_T^3} \right) m v_T^2 D_{ij}\left(\vec{c}\,\right) \phi\left(c\right) \, ,
\end{split}
\end{equation}
and recast Eq. \eqref{Icolltrans3} into the form
\begin{equation}
\label{eq:dimlessDijIcoll}
\int d \vec{V}_1 D_{ij}\left(\vec{V}_1\right) J^{(1)}\left(f^{(0)}, \gamma_{ij}\right)  
= 4 \eta   v_T  n  \sigma^2 \Omega_{\eta} \, ,  
\end{equation}
where we substitute $\gamma_0=-\eta/nT$  and where $\Omega_{\eta}$ is a numerical coefficient defined by 
\begin{multline}
\label{eq:def_Omegaeta_gen}
\Omega_{\eta} \equiv \int d \vec{c}_1 \int d \vec{c}_2 \int d\vec{e} 
        \Theta\left(-\vec{c}_{12} \cdot \vec{e}\,\right) \left|\vec{c}_{12} \cdot \vec{e}\,\right|  
        \tilde{f}^{(0)}\left(c_1\right)\\
\times \phi\left(c_2\right) D_{ij}\left(\vec{c}_2\right)\Delta \left[D_{ij}\left(\vec{c}_1\right) + D_{ij}\left(\vec{c}_2\right) \right] \, . 
\end{multline}
The coefficient $\Omega_\eta$ may be expressed in terms of the dissipation parameter $\delta^\prime$ and the second Sonine coefficient $a_2$ (see Appendix):
\begin{equation}
\label{eq:Omegaeta_vis}
\Omega_{\eta} =  -\left( w_0 + \delta^{\prime} w_1 - \delta^{\prime \, 2} w_2 \right)
\end{equation}
with 
\begin{equation}
\label{eq:Omega_www}
  \begin{split}
    w_0 =& 4  \sqrt{2 \pi} \left(1-\frac{1}{32}\,a_2 \right)\\
    w_1 =& \omega_0 \left(\frac{1}{15}-\frac{1}{500}\,a_2 \right) \\
    w_2 =& \omega_1 \left(\frac{97}{165}-\frac{679}{44000}\,a_2 \right)\, . 
  \end{split}
\end{equation}
Substituting Eqs. (\ref{eq:Dij_Cij},\ref{eq:dimlessDijIcoll},\ref{eq:Omegaeta_vis}) into Eq. \eqref{eqforeta} and using  Eq. \eqref{zetavisco} for $\zeta^{(0)}$ we obtain an equation for the coefficient of viscosity $\eta$:
\begin{multline}
\label{eqetavisco}
\left(\omega_0  \delta^{\prime} - \omega_2 \delta^{\prime \, 2} \right) T\frac{\partial \eta}{\partial T} \\ 
= \frac35\left(w_0 +w_1 \delta^{\prime} -w_2 \delta^{\prime \, 2} \right)\eta
-\frac32 \frac{1}{\sigma^2} \sqrt{\frac{mT}{2}} \, . 
\end{multline}
We seek the solution as an expansion in terms of $\delta^{\prime}$:
\begin{equation}
\label{etexpvisco}
\eta = \eta_0\left(1 + \delta^{\prime} \tilde{\eta}_1 + \delta^{\prime \, 2} \tilde{\eta}_2 + \dots \right)\,.
\end{equation}
The solution  in zeroth order
\begin{equation}
\label{etaelastic}
\eta_0 = \frac{5}{16 \sigma^2} \sqrt{\frac{mT}{\pi}}
\end{equation}
is the viscosity coefficient for a gas of elastic particles (Enskog viscosity), while the coefficients $\tilde{\eta}_1$ and $\tilde{\eta}_2$ account for the dissipative properties of viscoelastic particles. With Eq. \eqref{eq:derdelprime} the temperature derivative of the viscosity coefficient reads
\begin{equation}
\label{etaTdervisc}
T\frac{\partial \eta}{\partial T}=\eta_0 \left(\frac12 +\frac35 \delta^{\prime} \tilde{\eta}_1 +\frac{7}{10} \delta^{\prime \, 2} \tilde{\eta}_2 +\dots \right) \, .
\end{equation}
We substitute Eqs. (\ref{etexpvisco},\ref{etaTdervisc})  into Eq. \eqref{eqetavisco}, express $a_2$ in terms of $\delta^{\prime}$ according to Eq. \eqref{a2adiab} and collect terms of the same order in $\delta^{\prime}$. This yields the equations for  $\tilde{\eta}_1$,  $ \tilde{\eta}_2$, etc., whose solutions  read
\begin{equation}
\label{eta1eta2}
\begin{split}
\tilde{\eta}_1  = & \frac{359}{3840} \frac{\sqrt{2 \pi}}{\pi} \omega_0 \approx 0.483\\
\tilde{\eta}_2  = & \frac{41881}{2304000}\frac{\omega_0^2}{\pi} 
-\frac{567}{28160} \frac{\omega_1 \sqrt{2 \pi} }{\pi}  \approx 0.094 \, ,
\end{split}
\end{equation}
with $\omega_{0/1}$ given by Eq. \eqref{eq:omegadef}.

Thus, we arrive at the final expression for the viscosity coefficient for a granular gas of viscoelastic particles:
\begin{equation}
\label{eq:eta_fin}
\eta = \frac{5}{16 \sigma^2} \sqrt{\frac{mT}{\pi}} \left(1 + 0.483\, \delta^{\prime} + 0.094\, \delta^{\prime \, 2} + \dots \right)\,.
\end{equation}
In contrast to granular gases of simplified particles ($\varepsilon=\mbox{const.}$), where $\eta \propto \sqrt{T}$, for a gas of viscoelastic particles there is an additional temperature dependence due to the time-dependent coefficient $\delta^{\prime}$.

\section{Coefficient of thermal conductivity and the coefficient $\mu$}

To find coefficients $\kappa$ and $\mu$ we need the coefficients $\vec{\alpha}$ and $\vec{\beta}$ which are the solutions of Eqs. (\ref{eqsalbega1},\ref{eqsalbega2}). The functions $\vec{A}$ and $\vec{B}$ in the right-hand sides may be found from Eqs. (\ref{coefABCvisco1},\ref{coefABCvisco2}):
\begin{multline}
\label{eq:A_B}
\vec{A}  =  - \frac{1}{T} \vec{S} \left( \vec{V} \right) \left[ 1 + a_2 \left(S_2\left(c^2\right) +1-c^2 \right) \right] f_M \\
- \vec{V} S_2\left(c^2\right) \left[\frac{a_{21}}{10}\delta^{\prime} + \frac{a_{22}}{5}\delta^{\prime \, 2} \right]f_M
\end{multline}
\begin{equation}
\vec{B}  =  \frac{a_2}{T} \vec{S} \left( \vec{V} \right) f_M \, . 
\end{equation}
Keeping only leading terms with respect to the gradients and $a_2$, we choose $\vec{\alpha}$ in the form
\begin{equation}
\label{alpfirst}
\vec{\alpha} =-\frac{\alpha_1}{T} \vec{S} \left( \vec{V} \right) f_M(V) \,, 
\end{equation}
with $\alpha_1$ being the velocity independent coefficient. This Ansatz for $\vec{\alpha}$ yields the coefficient of thermal conductivity,
\begin{multline}
  \kappa = - \frac{1}{3T} \int d \vec{V} \vec{S}\left( \vec{V} \right) \cdot \vec{\alpha} \left( \vec{V} \right) \\
 =  \frac{\alpha_1}{3T^2} \int d \vec{V} \left( \frac{mV^2}{2} - \frac52 T \right)^2 V^2 f_M \\
             = \frac52 \frac{nT}{m} \alpha_1 \,,
\label{kapfin}
\end{multline}
which implies 
\begin{equation}
\label{eq:alpha1}
\alpha_1=\frac{2m}{5nT} \kappa\,.
\end{equation}
Multiplying Eq. \eqref{eqsalbega1} for $\vec{\alpha}$ by $\vec{S} \left.\left( \vec{V}_1 \right)\right/T$, integrating over $\vec{V}_1$ and  using Eq. \eqref{kapSmuSfin} for  $\kappa$ we obtain
\begin{equation}
\label{eqalphvisco}
3 \frac{\partial}{\partial T }\zeta^{(0)} \kappa T 
=\frac{1}{T} \! \int d \vec{V}_1 \vec{S} \cdot \vec{A}   
 -\frac{1}{T} \! \int d \vec{V}_1 \vec{S} \cdot J^{(1)}\left(f^{(0)}, \vec{\alpha}\right) \, .
\end{equation}
To evaluate the first term in the right-hand side  we use Eq. \eqref{1qdef} for $\vec{S}\left(\vec{V}\right)$ and Eq. \eqref{eq:A_B} for $\vec{A}\left(\vec{V}\right)$:
\begin{multline} 
\label{eq:SA_dV} 
-\frac{1}{T} \int d \vec{V} \vec{S} \left( \vec{V} \right) \cdot \vec{A} \left( \vec{V} \right) =  \\ 
= v_T^2 \int d \vec{V} \left( c^2 - \frac52  \right)^2 c ^2 
        \left[ 1 + a_2 \left(S_2\left(c^2\right) +1-c^2 \right) \right] f_M \\
+ v_T^2  \int d \vec{V} \left( c^2 - \frac52  \right) c^2 
        \left[\frac{a_{21}}{10}\delta^{\prime} +  \frac{a_{22}}{5}\delta^{\prime \, 2} \right] 
         S_2\left(c^2\right) f_M \,.
\end{multline} 
With the Maxwell distribution Eq. \eqref{eq:Maxwdistr} the first term in the last equation (\ref{eq:SA_dV}) reads
\begin{multline}
\label{eq:2ChapEns8} 
4 \pi n v_T^2  \left\{ 
\int_0^{\infty}c^4\phi(c)\left(c^2- \frac{5}{2}\right)^2 dc \right.\\
\left. +a_2 \int_0^{\infty}c^4\phi(c)\left(c^2- \frac{5}{2} \right)^2
\left[\frac{c^4}{2}-\frac{7c^2}{2}+\frac{23}{8} \right]dc \right\}\\
=\frac{15}{4}n v_T^2 +\frac{15}{2}n v_T^2 a_2 = \frac{15}{2} \frac{nT}{m}\left(1+2a_2\right) \, ,
\end{multline}
where integration over the angles has been performed and the integral  
\begin{equation}
\int_0^{\infty} \exp\left(-x^2\right) x^{2k} dx = \frac{(2k-1)!!}{2^{k+1}} \sqrt{\pi}
\end{equation}
was used. Very similar calculations give the second term in Eq. \eqref{eq:SA_dV}: 
\begin{equation}
\label{eq:2ChapEns81}
\frac{15}{4}n v_T^2
\left(\frac{a_{21}}{10}\delta^{\prime} +  \frac{a_{22}}{5}\delta^{\prime \, 2} \right) \, . 
\end{equation}
Summing up Eqs. \eqref{eq:2ChapEns8} and \eqref{eq:2ChapEns81} we obtain the first term in the right-hand side of \eqref{eqalphvisco}:
\begin{equation}
\label{SAvisco} 
\frac{1}{T} \int d \vec{V}_1 \vec{S}  \cdot \vec{A}   
=-\frac{15}{2}\frac{nT}{m} \left( 1+\frac{21}{10}a_{21} \delta^{\prime} +\frac{11}{5}a_{22} \delta^{\prime \, 2} \right)\,.
\end{equation}
The second  term in the right-hand side of Eq. \eqref{eqalphvisco} may be again written using the basic property of the collision integral [see Eq. \eqref{Icolltrans1}]:
\begin{multline}
\label{Icolltrankap}
-\frac{1}{T} \int d \vec{V}_1 \vec{S} \left( \vec{V}_1 \right) \cdot J^{(1)}\left(f^{(0)}, \vec{\alpha} \right) =\\
= \frac{\sigma^2}{T}  \int d \vec{V}_1 \int d \vec{V}_2 f^{(0)}\left(\vec{V}_1\right)\vec{\alpha} \left(\vec{V}_2\right) \cdot \int d\vec{e}\, \Theta\left(-\vec{V}_{12} \cdot \vec{e}\right)\\
\times \left|\vec{V}_{12} \cdot \vec{e}\,\right| \Delta \left[\vec{S} \left( \vec{V}_1 \right) + \vec{S} \left( \vec{V}_2 \right) \right]\,. 
\end{multline}
Using the dimensionless variables
\begin{equation}
\label{eq:S_alpha_dimless}
\begin{split}
\vec{S} \left( \vec{V} \right)      & =  v_T T \vec{S} \left( \vec{c} \right) = v_T T \left(c^2 - \frac52 \right) \vec{c} \\
\vec{\alpha} \left( \vec{V} \right) & = -\alpha_1 \, v_T \left( \frac{n}{\vec{v}_T^{\,3}} \right) \phi(c) \vec{S} \left( \vec{c}\, \right) 
\end{split}
\end{equation}
and Eq. (\ref{eq:alpha1}) for $\alpha_1$, we recast the last equation into the form
\begin{equation}
\label{Icolltrans4}
-\frac{1}{T}  \int d \vec{V} \vec{S} \left( \vec{V} \right) \cdot J^{(1)}\left(f^{(0)}, \vec{\alpha} \right)
=-\frac45 \kappa v_T n \sigma^2 \Omega_{\kappa}\,. 
\end{equation}
The coefficient $\Omega_{\kappa}$ is defined by
\begin{multline}
\label{eq:def_Omegakap_gen}
\Omega_{\kappa} \equiv \int d \vec{c}_1 \int d \vec{c}_2 \int d\vec{e} \Theta\left(-\vec{c}_{12} \cdot \vec{e}\,\right)\left|\vec{c}_{12} \cdot \vec{e}\right| \tilde{f}^{(0)}\left(c_1\right)\\
\times \phi\left(c_2\right) \vec{S} \left( \vec{c}_2  \right)  \cdot  \Delta \left[ \vec{S} \left( \vec{c}_1 \right) + \vec{S} \left( \vec{c}_2 \right) \right] \, . 
\end{multline}
This coefficient reads (see Appendix)
\begin{equation}
\label{eq:Omegakappa}
\Omega_{\kappa}  =  -\left( u_0 + \delta^{\prime} u_1 - \delta^{\prime \, 2} u_2 \right)
\end{equation}
where 
\begin{equation}
\label{eq:u0u1u2_kappa_vis}
\begin{split}
u_0  = & 4\sqrt{2 \pi} \left(1+\frac{1}{32}\,a_2 \right) \\
u_1  = & \omega_0 \left( \frac{17}{5}- \frac{9}{500}\,a_2 \right) \\
u_2  = & \omega_1 \left( \frac{1817}{440}- \frac{1113}{352000}\, a_2 \right) \, .  
\end{split}
\end{equation}        
Substituting Eqs. (\ref{SAvisco},\ref{Icolltrans4},\ref{eq:Omegakappa}) into Eq. \eqref{eqalphvisco} and using Eq. \eqref{zetavisco} for $\zeta^{(0)}$ we arrive at 
\begin{multline}
\label{eqforkapvisco}
T \frac{\partial }{\partial T}  \kappa \, T^{3/2} \left( \omega_0 \delta^{\prime}
  -\omega_2 \delta^{\prime \, 2}+\dots \right)\\
= \frac25  \kappa T^{3/2} \left(u_0+\delta^{\prime} u_1 - \delta^{\prime \, 2} u_2 +\dots \right)\\
-\frac{15}{4} \frac{T^{3/2}}{\sigma^2 } \sqrt{\frac{T}{2m}} 
\left( 1+\frac{21}{10}a_{21} \delta^{\prime} +\frac{11}{5}a_{22} \delta^{\prime \, 2} \right)  \,.
\end{multline}
We solve this equation with the Ansatz
\begin{equation}
\label{kapexpvisco}
\kappa = \kappa_0\left(1 + \delta^{\prime} \tilde{\kappa}_1 + \delta^{\prime \, 2} \tilde{\kappa}_2 +\dots \right)\,,
\end{equation}
where 
\begin{equation}
\label{kapelastic}
\kappa_0=\frac{75}{64 \sigma^2 } \sqrt{\frac{T}{\pi m}}
\end{equation}
is the Enskog thermal conductivity for a gas of elastic particles. Substituting Eq. \eqref{kapexpvisco} into Eq. \eqref{eqforkapvisco} and equating terms of the same order in $\delta^{\prime}$ we obtain the coefficients
\begin{equation}
\label{kappa12visco}
\begin{split}
\tilde{\kappa}_1 & = \frac{487}{6400} \frac{ \sqrt{2 \pi}\omega_0 }{\pi} \approx 0.393\\
\tilde{\kappa}_2 & = \frac{1}{\pi} \left(\frac{2872113}{51200000}\omega_0^2 +\frac{78939}{140800} \sqrt{2 \pi} \omega_1 \right) \approx 4.904 \, . 
\end{split}
\end{equation}
Hence, the coefficient of thermal conductivity for a granular gas of viscoelastic particles reads in adiabatic approximation 
\begin{equation}
\label{eq:kappa_fin}
\kappa=\frac{75}{64 \sigma^2 } \sqrt{\frac{T}{\pi m}} 
\left(1 + 0.393 \delta^{\prime} + 4.904 \delta^{\prime \, 2} + \dots \right)\,.
\end{equation}
Similar to the viscosity coefficient, the coefficient of thermal conductivity of a granular gas of viscoelastic particles reveals an additional temperature dependence as compared with gases of simplified particles where $\varepsilon=\mbox{const.}$

The evaluation of the coefficient $\mu$ may be performed in the same way as $\kappa$, i.e. choosing $\vec{\beta}$ in the form  
\begin{equation}
\label{betfirst}
\vec{\beta} =-\frac{\beta_1}{T} \vec{S} \left( \vec{V} \right) f_M(V) \,, 
\end{equation}
with the velocity independent coefficient $\vec{\beta}$. The only difference is that the expansion of $\mu$ in terms of the dissipative parameter $\delta^{\prime}$ lacks the term in zeroth order since $\mu$ vanishes in the elastic limit. Since the calculations are completely analogous to that for $\kappa$, we present here only the final result: 
\begin{equation}
\label{eq:mu_fin}
\mu = \frac{\kappa_0T}{n} \left(\delta^{\prime} \tilde{\mu}_1 + \delta^{\prime \, 2} \tilde{\mu}_2
+\dots \right) 
\end{equation} 
with 
\begin{equation}
\label{eq:mu_12}
\begin{split}
\tilde{\mu}_1  = & \frac{19}{80} \frac{\omega_0 \sqrt{2 \pi} }{\pi} \approx  1.229\\
\tilde{\mu}_2  = & \frac{1}{\pi} \left( \frac{58813}{640000} \omega_0^2 - \frac{1}{40} \sqrt{2 \pi}\omega_1 \right) \approx 1.415 \, .
\end{split}
\end{equation}
Thus, the coefficient $\mu$ reads in adiabatic approximation
\begin{equation}
\label{eq:mu_fin1}
\mu = \frac{\kappa_0T}{n} \left( 1.229\, \delta^{\prime} + 1.415\, \delta^{\prime \, 2} +\dots \right) 
\end{equation} 
Finally, using the coefficients 
\begin{equation}
\label{albegakamuet}
\alpha_1=\frac{2m}{5nT} \kappa \, ,\qquad
\beta_1=\frac{2m}{5T^2} \mu \, , \qquad
\gamma_0= - \frac{1}{nT} \eta \, ,
\end{equation} 
(the result for $\beta_1$ may be derived analogously as for $\alpha_1$) in the relations (\ref{alpfirst},\ref{betfirst},\ref{1gamfirst}) for $\vec{\alpha}$, $\vec{\beta}$, $\gamma_{ij}$ we obtain an expression for the first-order  distribution function $f^{(1)}$, which depends linearly on the field gradients according to Eq. (\ref{f1form}). 

\section{Two-dimensional Granular Gas}
So far we have restricted ourselves to three-dimensional systems although the calculations are identical for general dimension $d$. Of particular interest is the case $d=2$. Since  molecular dynamics simulations are frequently performed for two-dimensional systems, in this section we present the results for two-dimensional gases of viscoelastic particles. We wish to stress that these systems are, in fact, quasi two-dimensional, since we still use the coefficient of restitution Eq. \eqref{epsilon} for colliding spheres. Hence, we assume that the motion of the spherical particles of the gas is restricted to a two-dimensional surface. 

The hydrodynamic equations for two-dimensional gases have the form
\begin{equation}
\label{eq:hydroeq_genD}
\begin{split}
    \frac{\partial n}{\partial t} + \vec{\nabla} \cdot \left( n \vec{u} \right) & =0 \\
    \frac{\partial \vec{u}}{\partial t} + \vec{u} \cdot \vec{\nabla} \vec{u} + \left( n m \right)^{-1} \vec{\nabla} \cdot \hat {P} &=0 \\
    \frac{\partial T}{\partial t} +  \vec{u} \cdot \vec{\nabla} T + \frac{1}{n} \left( P_{ij} \nabla_j u_i + \vec{\nabla} \cdot \vec{q} \right) + \zeta \, T &=0  \,,  
\end{split}
\end{equation}
with the pressure tensor and the heat flux
\begin{equation}
\label{eq:Pres_q_genD}
\begin{split}
    P_{ij}  & = nT \delta_{ij} - \eta \left( \nabla_i u_j +\nabla_j u_i - \delta_{ij} \vec{\nabla} \cdot \vec{u} \right)  \\
    \vec{q} & = -\kappa \vec{\nabla} T - \mu \vec{\nabla} n   \,.  
\end{split}
\end{equation}
Correspondingly, the transport coefficients read
\begin{equation}
\label{eq:eta_2D_visco}
\eta = \eta_0 \left(1 + \delta^{\prime} \tilde{\eta}_1 + \delta^{\prime \, 2} \tilde{\eta}_2  
       + \cdots \right) \, , 
\end{equation}
where $\delta^{\, \prime} \left(t\right) = \delta \left[ 2T \left( t \right) / T_0  \right]^{1/10}$ and 
\begin{eqnarray}
\label{eq:eta012_2D_visco}
             \eta_0 & = & \frac{1}{2 \sigma}  \sqrt{\frac{mT}{\pi}} \nonumber \\
     \tilde{\eta}_1 & = & \frac{29}{640} \frac{\sqrt{2 \pi}}{\pi} \omega_0 \approx 0.234 \\
     \tilde{\eta}_2 & = & \frac{111}{160 000}\frac{\omega_0^2}{\pi} 
                   +\frac{569}{14080} \frac{\omega_1 \sqrt{2 \pi} }{\pi}  \approx 0.308 \,. \nonumber 
\end{eqnarray} 
Similarly, 
\begin{equation}
\label{eq:kappa_2D_visco}
\kappa = \kappa_0\left(1 + \delta^{\prime} \tilde{\kappa}_1 + \delta^{\prime \, 2} \tilde{\kappa}_2 +\cdots \right)
\end{equation}
with  
\begin{eqnarray} 
\label{eq:kappa012_2D_visco}
     \kappa_0         & = & \frac{2}{\sigma } \sqrt{\frac{T}{ \pi m }} \nonumber \\
     \tilde{\kappa}_1 & = &-\frac{433}{3200} \frac{ \sqrt{2 \pi}\omega_0 }{\pi} \approx -0.700\\
     \tilde{\kappa}_2 & = & \frac{1}{\pi} \left( \frac{1749573}{12800000}\omega_0^2 +\frac{95619}{70400} \sqrt{2 \pi} \omega_1 \right) \approx 11.89\,. \nonumber 
\end{eqnarray} 
and 
\begin{eqnarray} 
\label{eq:MU12_2D_visco}
       \mu   & = & \frac{\kappa_0T}{n} \left(\delta^{\prime} \tilde{\mu}_1 + \delta^{\prime \, 2} \tilde{\mu}_2 +\cdots \right) \nonumber \\
     \tilde{\mu}_1 & = & \frac{7}{20} \frac{\omega_0 \sqrt{2 \pi} }{\pi}  \approx 1.811 \\
     \tilde{\mu}_2 & = & \frac{1}{\pi} \left( -\frac{7411}{80 000} \omega_0^2 + \frac{7}{40} \sqrt{2 \pi}\omega_1 \right) \approx 0.056\,. \nonumber 
\end{eqnarray} 
The cooling rate $\zeta^{(0)}$ reads for a two-dimensional gas of viscoelastic particles
\begin{eqnarray} 
\label{eq:zeta0_2D_visco}
  \zeta^{(0)}    & = & \sqrt{\frac{2T}{m}} n \sigma   \left(\frac12 \omega_0 \delta^{\prime} - \tilde{\omega}_2 \delta^{\prime \, 2} + \cdots  \right) \nonumber \\
    \tilde{\omega}_2 & = & \frac12 \omega_1 + \frac{9 \sqrt{2 \pi}}{500 \pi} \omega_0^2 \approx 5.246\,.
\end{eqnarray} 
The numerical constants $\omega_{0/1}$ are given by Eq. \eqref{eq:omegadef}.

\section{Summary}

We have derived the hydrodynamics of granular gases of viscoelastic particles. Collisions of viscoelastic particles are characterized by an impact velocity dependent coefficient of restitution. We have used the Chapman-Enskog approach together with an adiabatic approximation for the velocity distribution function, which assumes that the shape of the velocity distribution function follows adiabatically the decaying temperature. We have compared the numerical solutions for temperature $T(t)$ and for the second Sonine coefficient $a_2(t)$ with the corresponding adiabatic approximations and have found good agreement up to intermediate dissipation. To derive the hydrodynamic equations and transport coefficients from the  Boltzmann equation we have modified the standard scheme to account for the time dependence of the basic solution. This modification takes into account the time dependence not only due to the thermal velocity, as for $\varepsilon = \mbox{const.}$, but also  due to the evolution of the shape of the distribution function as given for gases of viscoelastic particles. 

Transport coefficients and the cooling coefficient for dilute granular gases of viscoelastic particles read
\begin{eqnarray}
\label{etexpvisco1}
\eta &=& \eta_0\left(1 + \delta^{\prime} \tilde{\eta}_1 + \delta^{\prime \, 2} \tilde{\eta}_2 + \dots \right) \\
\label{kapexpvisco1}
\kappa &=& \kappa_0\left(1 + \delta^{\prime} \tilde{\kappa}_1 + \delta^{\prime \, 2} \tilde{\kappa}_2 +\dots \right) \\
\label{1eq:mu_fin1}
\mu &=& \frac{\kappa_0T}{n} \left( \delta^{\prime}  \tilde{\mu}_1+ \delta^{\prime \, 2} \tilde{\mu}_2  + \dots \right)  \\
\label{eq:zetax}
\zeta^{(0)} & = & \frac{nT}{\eta_0} \left( \delta^{\prime}  \tilde{\zeta}_1+ \delta^{\prime \, 2} \tilde{\zeta}_2  + \dots \right)  
\end{eqnarray}
where $\eta_0$ and $\kappa_0$ are the Enskog values for the viscosity and the coefficient of thermal conductivity and $\delta^{\prime}$ is the time dependent dissipative parameter. The numerical coefficients $\tilde{\eta}_{1/2}$, $\tilde{\kappa}_{1/2}$, $\tilde{\mu}_{1/2}$, and $\tilde{\zeta}_{1/2}$ are given in Table \ref{tab:1}.

\begin{table}[htbp]
  \begin{center}
    \caption{Numerical coefficients for Eqs. (\ref{etexpvisco1}-\ref{eq:zetax})}
  \begin{tabular}{lll|lll}
\hline\hline
& 3$d$~~~~~~~ & 2$d$~~~~~~ & &  3$d$~~~~~~~ & 2$d$\\ \hline
$\tilde{\eta}_1$~~~ & 0.483 & 0.234 &~
$\tilde{\eta}_2$~~~ & 0.094 & 0.309)\\
$\tilde{\kappa}_1$ & 0.393 & -0.700 &~
$\tilde{\kappa}_2$ & 4.904 &  11.893\\
$\tilde{\mu}_1$ & 1.229 & 1.811 &~
$\tilde{\mu}_2$ & 1.415 & 0.056\\
$\tilde{\zeta}_1$ & 1.078 & 1.294 &~ 
$\tilde{\zeta}_2$ & -1.644 & -2.093\\
\hline\hline
  \end{tabular}
    \label{tab:1}
  \end{center}
\end{table}

The dependence on temperature and, therefore, on time of the kinetic coefficients Eqs. \eqref{etexpvisco1}-\eqref{eq:zetax} differs significantly from the time dependence of the corresponding coefficients for granular gases of particles which collide with a simplified collisional model $\varepsilon = \mbox{const.}$  This leads to the serious consequences for the global behavior of force-free granular gases  \cite{BrilliantovPoeschel:2001roy}.   

Under mild preconditions the presented formalism for the derivation of the hydrodynamic equations and the transport coefficients may be applied also to gases of particles whose collision is described by a different impact velocity dependence than given for viscoelastic particles.

\appendix
\section{Derivation of the coefficients $\Omega_\eta$ and $\Omega_\kappa$}
For the evaluation of the numerical coefficient Eq. \eqref{eq:def_Omegaeta_gen}, defined by
\begin{multline}
\label{eq:def_Omegaeta_genAPP}
\Omega_{\eta} \equiv \int d \vec{c}_1 \int d \vec{c}_2 \int d\vec{e}\, 
        \Theta\left(-\vec{c}_{12} \cdot \vec{e}\,\right) \left|\vec{c}_{12} \cdot \vec{e}\,\right|  
        \tilde{f}^{(0)}\left(c_1\right)\\
\times \phi\left(c_2\right) D_{ij}\left(\vec{c}_2\right)\Delta \left[D_{ij}\left(\vec{c}_1\right) + D_{ij}\left(\vec{c}_2\right) \right] \, . 
\end{multline}
we need the factor 
\begin{multline} 
\label{eq:DijdelDij}
D_{ij}\left(\vec{c}_2\right) \Delta \left[D_{ij}\left(\vec{c}_1\right) + D_{ij}\left(\vec{c}_2\right) \right]\\
=\left( \vec{c}^{\,\prime}_1 \cdot \vec{c}_2 \right)^2 + \left( \vec{c}^{\,\prime}_2 \cdot \vec{c}_2 \right)^2 - \left( \vec{c}_1 \cdot \vec{c}_2 \right)^2 - \left( \vec{c}_2 \cdot \vec{c}_2 \right)^2\\
-\frac13 c_2^2 \left( c^{\prime\,2}_1 + c^{\prime\,2}_2 -c_1^2-c_2^2  \right)
\end{multline}
Similarly, for the coefficient
\begin{multline}
\label{eq:def_Omegakap_genAPP}
\Omega_{\kappa} \equiv \int d \vec{c}_1 \int d \vec{c}_2 \int d\vec{e}\, \Theta\left(-\vec{c}_{12} \cdot \vec{e}\,\right)\left|\vec{c}_{12} \cdot \vec{e}\,\right| \tilde{f}^{(0)}\left(c_1\right)\\
\times \phi\left(c_2\right) \vec{S} \left( \vec{c}_2  \right)  \cdot  \Delta \left[ \vec{S} \left( \vec{c}_1 \right) + \vec{S} \left( \vec{c}_2 \right) \right] \,, 
\end{multline}
given by Eq. \eqref{eq:def_Omegakap_gen} we need 
\begin{multline} 
\label{eq:SdelS}
\vec{S} \left( \vec{c}_2 \right) \cdot \Delta \left[\vec{S} \left( \vec{c}_1 \right) + \vec{S} \left( \vec{c}_2 \right) \right] 
=\left( c_2^2-\frac52 \right) \left[ \left( \vec{c}^{\,\prime}_1 \cdot \vec{c}_2 \right) \left(c^{\,\prime}_1 \right)^2   \right.\\
+\left. \left( \vec{c}^{\,\prime}_2 \cdot \vec{c}_2 \right)  \left(c^{\,\prime}_2 \right)^2 - \left( \vec{c}_1 \cdot \vec{c}_2 \right) c_1^2 - \left( \vec{c}_2 \cdot \vec{c}_2 \right) c_2^2 \right] \, .
\end{multline}
With \eqref{eq:DijdelDij} and \eqref{eq:SdelS} we write for the coefficient $ \Omega_{\eta}$
\begin{multline}
\label{eq:def_Omegaetael}
\Omega_{\eta} =  \int d \vec{c}_1 \int d \vec{c}_2 \int d\vec{e}\, 
\Theta\left(-\vec{c}_{12} \cdot \vec{e}\,\right) \left|\vec{c}_{12} \cdot \vec{e}\,\right|  \tilde{f}^{(0)}(c_1)\\
\times \phi\left(c_2\right)\left[ \left(\vec{c}^{\,\prime}_1 \cdot \vec{c}_2 \right)^2 + \left( \vec{c}^{\,\prime}_2 \cdot \vec{c}_2 \right)^2  - \left( \vec{c}_1 \cdot \vec{c}_2 \right)^2 \right.\\
\left. - \left( \vec{c}_2 \cdot \vec{c}_2 \right)^2 -\frac13 c_2^2 \left( c^{\prime\,2}_1 + c^{\prime\,2}_2 -c_1^2-c_2^2  \right) \right] \, . 
\end{multline}
and for the coefficient $ \Omega_{\kappa}$, 
\begin{multline} 
\label{eq:def_Omegakap_genX}
\Omega_{\kappa} = \int d \vec{c}_1 \int d \vec{c}_2 \int d\vec{e}\, 
\Theta\left(-\vec{c}_{12} \cdot \vec{e}\,\right) \left|\vec{c}_{12} \cdot \vec{e}\,\right| \tilde{f}^{(0)}\left(c_1\right)\\
\times  \phi_2\left(c_2\right) \left( c_2^2-\frac52 \right) 
\left[ \left( \vec{c}^{\,\prime}_1 \cdot \vec{c}_2 \right)  \left(c^\prime_1 \right)^2  + \left( \vec{c}^{\,\prime}_2 \cdot \vec{c}_2 \right)  \left(c^\prime_2 \right)^2 \right.\\ 
\left. - \left( \vec{c}_1 \cdot \vec{c}_2 \right) c_1^2 - \left( \vec{c}_2 \cdot \vec{c}_2 \right) c_2^2 \right] \,. 
\end{multline} 
The pre-collision velocities $\vec{c}_1$, $\vec{c}_2$ as well as after-collision velocities $\vec{c}^{\,\prime}_1$, $\vec{c}^{\,\prime}_2$ can be expressed in terms of the center of mass velocity $\vec{C}=\left(\vec{c}_1+\vec{c}_2\right)/2$ and the relative velocity $\vec{c}_{12}=\vec{c}_1-\vec{c}_2$ before the collision:
\begin{equation}
\label{1directcollviaCc}
\begin{split}
\vec{c}_1&=\vec{C}+\frac12 \vec{c}_{12}\\
\vec{c}_2&=\vec{C}-\frac12 \vec{c}_{12}\\
\vec{c}^{\,\prime}_1 &=\vec{C}+\frac12 \vec{c}_{12} - \frac{1}{2}\left(1+\varepsilon\right)\left(\vec{c}_{12} \cdot \vec{e}\,\right)\vec{e} \\
\vec{c}^{\,\prime}_2 &=\vec{C}-\frac12 \vec{c}_{12} + \frac{1}{2}\left(1+\varepsilon\right) \left(\vec{c}_{12} \cdot \vec{e}\,\right)\vec{e}\, .
\end{split}
\end{equation}
The coefficient of restitution is expressed in terms of the relative velocity $\vec{c}_{12}$ by
\begin{equation}
\label{epsc12}
\varepsilon= 1-C_1\delta^{\prime}\left(t\right) \left|\vec{c}_{12} \cdot \vec{e}\, \right|^{1/5}+
\frac{3}{5}C_1^2 \,\delta^{\prime \, 2}\left(t\right) \left|\vec{c}_{12} \cdot 
\vec{e}\, \right|^{2/5} + \dots 
\end{equation} 
The second Sonine polynomial in the distribution function
\begin{equation}
\label{eq:f0again}
\tilde{f}^{(0)}(c_1) =\phi(c_1) \left[ 1 +a_2 S_2\left(c_1^2\right) \right] \, , 
\end{equation} 
reads in terms of $\vec{C}$ and $\vec{c}_{12}$
\begin{multline}
\label{S2c1Cc12}
S_2\left(c_1^2\right)=\frac{C^4}{2}+\frac12\left(\vec{C} \cdot \vec{c}_{12}\right)^2
+\frac{1}{32}c_{12}^4 + C^2 \left(\vec{C} \cdot \vec{c}_{12}\right)\\
+\frac14 C^2 c_{12}^2 +\frac14c_{12}^2\left(\vec{C} \cdot \vec{c}_{12}\right) 
-\frac52 C^2 \\
-\frac52 \left(\vec{C} \cdot \vec{c}_{12}\right) -\frac58 c_{12}^2 +\frac{15}{8} \, . 
\end{multline}
If we replace all factors  in the integrands of Eqs. (\ref{eq:def_Omegaetael},\ref{eq:def_Omegakap_genX}) by the corresponding expressions in terms of $\vec{C}$ and $\vec{c}_{12}$ we observe that $\Omega_{\eta}$ and $ \Omega_{\kappa}$ may be written as a sum of integrals of the structure
\begin{multline}
\label{eq:basin_App}
J_{k,l,m,n,p,\alpha}= 
\int d \vec{c}_{12} \int d \vec{C} \int d\vec{e} \, 
\Theta\left(-\vec{c}_{12} \cdot \vec{e}\,\right) \left|\vec{c}_{12} \cdot \vec{e}\,\right|^{1+\alpha}\\ 
\phi\left(c_{12}\right) \phi(C)
C^k c_{12}^l \left(\vec{C} \cdot \vec{c}_{12} \right)^m \left(\vec{C} \cdot \vec{e}\, \right)^n 
\left(\vec{c}_{12} \cdot \vec{e}\, \right)^p \,.
\end{multline}
The solution of this integral in general dimension $d$ reads \cite{BrilliantovPoeschel:2000visc}
for $n=0$
\begin{multline}
\label{eq:bas_n0}
J_{k,l,m,0,p,\alpha}  = (-1)^p \left[ 1 + (-1)^m \right] 2^{l+m+p+\alpha+1} \Omega_d^{-1}\\
\times \, \beta_{p+\alpha+1} \beta_m \gamma_{k+m} \gamma_{l+m+p+\alpha+1}\, , 
\end{multline}
for $n=1$ 
\begin{multline}
\label{eq:bas_n1}
J_{k,l,m,1,p,\alpha}   = (-1)^{p+1} \left[ 1 + (-1)^{m+1} \right] 2^{l+m+p+\alpha+1} \Omega_d^{-1}\\ 
\times \,\beta_{p+\alpha+2} \beta_{m+1} \, \gamma_{k+m+1} \gamma_{l+m+p+\alpha+1}\, ,  
\end{multline}
and for $n=2$,  
\begin{multline}
  \label{eq:bas_n2}
  J_{k,l,m,2,p,\alpha} = (-1)^{p} \left[ 1 + (-1)^{m} \right] 2^{l+m+p+\alpha+1} \\
  \times  \left[ (d-1) \Omega_d \right]^{-1} \gamma_{k+m+2} \, \gamma_{l+m+p+\alpha+1} \\
  \times \left[  \left(d \beta_{p+\alpha+3} - \beta_{p+\alpha+1} \right)\beta_{m+2}  +  \left( \beta_{p+\alpha+1} -  \beta_{p+\alpha+3} \right) \beta_{m} \right]\,,  
\end{multline}
where 
\begin{equation}
\Omega_d  = \frac{2 \pi^{d/2}}{\Gamma \left(\frac{d}{2}\right)}  
\end{equation}
is the surface of a $d$-dimensional unit sphere. The coefficients $\beta_m$, $\gamma_m$ read 
\begin{equation}
  \begin{split}
    \beta_m  &= \pi^{(d-1)/2} \, \frac{ \Gamma \left( \frac{m+1}{2} \right) }{ \Gamma \left( \frac{m+d}{2} \right)}\\
    \gamma_m  &= 2^{-m/2} \, \, \frac{ \Gamma \left( \frac{m+d}{2} \right) }{\Gamma \left( \frac{d}{2} \right)}\,.    
  \end{split}
\end{equation} 

Following this procedure we obtain the desired coefficients as given by Eqs. \eqref{eq:Omegaeta_vis} and   \eqref{eq:Omegakappa}. 

The evaluation of the sums which lead to the factors $\Omega_\eta$ and $\Omega_{\kappa}$ is straightforward, however, very lengthy. They can be calculated by symbolic algebra \cite{PoeschelBrilliantovMaple:2003}. 


\end{document}